\documentclass[12pt,preprint]{aastex}
\usepackage{emulateapj5}
\usepackage{apjfonts}
\usepackage{epsfig}
\usepackage{natbib}

\newcommand{\chisq}{\ensuremath{\chi^2}}
\newcommand{\chisqr}{\ensuremath{\chi^2_r}}

\newcommand{\etal}{et al.}

\newcommand{\feii}{\ion{Fe}{2}}

\def\gtrsim{\mathrel{\hbox{\rlap{\hbox{\lower4pt\hbox{$\sim$}}}\hbox{\raise2pt\hbox{$>$}}}}}

\newcommand{\halpha}{H\ensuremath{\alpha}}

\newcommand{\hst}{\emph{HST}}

\newcommand{\kms}{km~s\ensuremath{^{-1}}}

\newcommand{\lf}{\ensuremath{L_{\rm{5100 \AA}}}}
\newcommand{\lum}{ergs s$^{-1}$}

\newcommand{\mli}{\ensuremath{\Upsilon_I}}

\newcommand{\mbh}{\ensuremath{M_\mathrm{BH}}}

\newcommand{\mgb}{\ion{Mg}{1}$b$}
\newcommand{\mlb}{\ensuremath{M_{\mathrm{BH}}-L_{\mathrm{bulge}}}}
\newcommand{\mgal}{\ensuremath{M_{\mathrm{BH}}-M_{\mathrm{bulge}}}}
\newcommand{\msigma}{\ensuremath{M_{\mathrm{BH}}-\sigmastar}}
\newcommand{\msun}{\ensuremath{M_{\odot}}}

\newcommand{\oii}{[\ion{O}{2}]}
\newcommand{\oiii}{[\ion{O}{3}]}

\newcommand{\sers}{S{\'e}rsic}

\newcommand{\sigmastar}{\ensuremath{\sigma_{\ast}}}
\newcommand{\sigmam}{\ensuremath{\sigma_{\ast {\rm m}}}}

\newcommand{\spitzer}{\emph{Spitzer}}

\def\lax{{$\mathrel{\hbox{\rlap{\hbox{\lower4pt\hbox{$\sim$}}}\hbox{$<$}}}$}}
\def\gax{{$\mathrel{\hbox{\rlap{\hbox{\lower4pt\hbox{$\sim$}}}\hbox{$>$}}}$}}

\slugcomment{To appear in {\it The Astrophysical Journal}.}
\shortauthors{GREENE, HO, \& BARTH}

\begin{document}

\title{Black Holes in Pseudobulges and Spheroidals: A Change in the Black Hole-Bulge Scaling Relations at Low Mass\altaffilmark{1}}

\stepcounter{footnote}

\author{Jenny E. Greene\altaffilmark{2}}
\affil{Department of Astrophysical Sciences, Princeton University, 
Princeton, NJ}

\author{Luis C. Ho}
\affil{The Observatories of the Carnegie Institution of Washington,
813 Santa Barbara St., Pasadena, CA 91101}

\author{Aaron J. Barth}
\affil{Department of Physics and Astronomy, University of California at 
Irvine, 4129 Frederick Reines Hall, Irvine, CA 92697}

\altaffiltext{1}{Based on
observations made with the NASA/ESA Hubble Space Telescope, obtained
at the Space Telescope Science Institute, which is operated by the
Association of Universities for Research in Astronomy, Inc., under
NASA contract NAS 5-26555. These observations are associated with
program GO-10596.}

\altaffiltext{2}{Hubble, Princeton-Carnegie Fellow}

\begin{abstract}

We investigate the relationship between black hole mass and host
galaxy properties for active galaxies with the lowest black hole
masses currently known in galaxy nuclei.  {\emph{Hubble Space
Telescope}} imaging confirms that the host galaxies have
correspondingly low luminosity; they are $\sim 1$ mag below $L^*$. In
terms of morphology, $\sim 60\%$ of the sample are disk-dominated, and
all of these are consistent with containing a bulge or (more likely)
pseudobulge, while the remainder are compact systems with no
discernible disk component.  In general the compact components of the
galaxies do not obey the fundamental plane of giant elliptical
galaxies and classical bulges, but rather are less centrally
concentrated at a given luminosity, much like spheroidal galaxies.
Our results strongly confirm that a classical bulge is not a
requirement for a nuclear black hole.  At the same time, the observed
ratio of black hole to bulge mass is nearly an order of magnitude
lower in this sample than that seen for classical bulges.  While the
\msigma\ relation appears to continue to low mass, it seems that black
hole-galaxy scaling relations do depend on galaxy structure.

\end{abstract}


\section{Introduction}

The discovery of a tight correlation of supermassive black hole (BH)
mass with the properties of the elliptical galaxies and classical
bulges in which they are found (e.g.,~the \msigma\ relation; Tremaine
\etal\ 2002) suggests that there is an underlying tight correspondence
between the growth of the two galaxy components.  Furthermore, the
extremely regular structural properties of elliptical galaxies and
spiral bulges (the fundamental plane; Dressler \etal\ 1987; Djorgovski
\& Davis 1987) imply very similar formation paths for the two
populations; apparently one aspect of that formation involves the
growth of a supermassive BH.  For instance, if elliptical galaxies
grow through hierarchical merging, perhaps BH accretion episodes are
instigated during violent merging (e.g.,~Hopkins \etal\ 2006), and
this accretion self-regulates the growth of the BH according to the
depth of the galaxy potential well, thus leading to the \msigma\
relation (e.g.,~Silk \& Rees 1998).  In principle, a useful test of
the BH-bulge connection is to look at the masses of BHs in galaxies
that have not developed classical bulges, to see whether the BH masses
in these galaxies have a similar or different relation to their
surrounding hosts.

Due to technical challenges, BH demographics in late-type galaxies are
virtually unconstrained.  Low-mass BHs are difficult to weigh because
they have a very small gravitational sphere of influence.  The only
late-type galaxies with stellar-dynamical constraints on the presence
of nuclear BHs are within the Local Group; the late-type spiral galaxy
M33 (Gebhardt \etal\ 2001) and the spheroidal galaxy NGC 205 (Valluri
\etal\ 2005) have upper limits on their BH masses significantly below
expectations based on the \msigma\ relation.  This is not necessarily
surprising; if BH growth is associated with bulge formation, then
perhaps we should not expect to find BHs in galaxies with little or no
bulges, or whose bulgelike central concentration formed through very
different mechanisms than did classical bulges.  Furthermore, low-mass
galaxies may have trouble retaining their BHs.  The anisotropic
gravitational radiation associated with an unequal-mass BH merger
event can impart a net velocity to the BH remnant that exceeds the
escape velocity of a dwarf galaxy (e.g.,~Merritt \etal\ 2004;
Volonteri 2007).  All theoretical suppositions aside, better
observational constraints are needed on the occupation fraction of BHs
in low-mass galaxies.  Without the luxury of dynamical constraints, we
are forced to use indirect tracers, in the form of nuclear activity,
to constrain the presence or absence of BHs in low-mass galaxies.

There are two well-studied low-mass galaxies with active nuclei.  NGC
4395 is an Sdm spiral galaxy only 4 Mpc away (Filippenko \& Sargent
1988; Filippenko \& Ho 2003) while POX 52 (Kunth \etal\ 1987) is a
spheroidal galaxy whose nature has been recently revisited based on
updated measurements (Barth \etal\ 2004; Thornton \etal\ 2008).
Both of these galaxies and the globular cluster G1 (Gebhardt \etal\
2002, 2005; Ulvestad \etal\ 2007) appear to obey the low-mass
extrapolation of the \msigma\ relation.  This intriguing fact inspired
Greene \& Ho (2004, 2007c) to search systematically for BHs with
masses $< 10^6$~\msun\ using the Sloan Digital Sky Survey (York \etal\
2000).  Follow-up spectroscopy revealed that, like their nearby
counterparts, they also obey the low-mass extrapolation of the
\msigma\ relation (Barth \etal\ 2005).  However, the physical
significance of \sigmastar\ for these systems is not immediately clear
in the absence of more detailed host galaxy structural information.
In this work we use \emph{Hubble Space Telescope} (\hst) Advanced
Camera for Surveys (ACS) imaging to constrain the structural
properties of the host galaxies of the 19 low-mass BHs from Greene \&
Ho (2004).

One of the particularly appealing aspects of this sample is that in
addition to low mass, the objects are radiating at high fractions of
their Eddington limit.  We thus have the opportunity to perform a
detailed study of the host galaxy environments of BHs that are
experiencing a significant growth episode, which is impossible for
more massive BHs that are no longer actively accreting locally
(e.g.,~Heckman \etal\ 2004; Greene \& Ho 2007b).  A number of
suggestions have been made that low-mass, high-Eddington ratio AGNs
have anomalously low BH masses compared to their surrounding galaxies,
based both on the linewidths of the narrow emission lines (e.g.,~Bian
\& Zhao 2004; Grupe \& Mathur 2004), and the circular velocity of the
host inferred from the integrated \ion{H}{1} linewidth (Ho \etal\
2008).  Although the Greene \& Ho systems appear to obey the low-mass
extrapolation of the \msigma\ relation (Barth \etal\ 2005), it will be
very instructive to examine the host galaxy luminosity and structure
as well.

A brief explanation is in order regarding our use of terminology.  A
principal goal of the study is to use the locations of our AGN host
galaxies on the fundamental plane to infer the nature and formation
histories of these galaxies. We therefore wish to be very careful that
our naming conventions reflect what we believe are physically
meaningful distinctions between different classes of objects.
Elliptical galaxies and the classical bulges that are found in the
centers of early-type disk galaxies both obey the fundamental plane
(e.g.,~Bender \etal\ 1992).  However, not all bulge components appear
to be built like small elliptical galaxies.  Particularly in late-type
spirals, bulge components increasingly depart from the structural and
stellar population properties that define elliptical galaxies and
begin to look more like disks.  These so-called pseudobulges are
characterized by a high degree of rotational support, exponential
profiles, ongoing star formation or starbursts, and nuclear bars or
spirals (see review in Kormendy \& Kennicutt 2004).  In general, the
properties of pseudobulges suggest that they have formed through
dissipative processes and have evolved secularly, over many disk
rotation periods.

At even lower galaxy mass, there is yet another sequence of stellar
systems, of which NGC 205 is a local example.  These objects are
small-scale, hot stellar systems, and thus are often called
``dwarf elliptical galaxies''.  However, based on their structural
properties (specifically, their position in the fundamental plane)
they are nothing like small elliptical galaxies.  There are real
elliptical galaxies at these luminosities, such as M32.  In contrast,
spheroidal galaxies, as we will call them, are clearly less centrally
concentrated than elliptical galaxies at the same luminosity
(e.g.,~Wirth \& Gallagher 1984; Kormendy 1985).  The alternate view,
that elliptical and spheroidal galaxies form a continuous sequence,
has been expressed in the literature, based predominantly on the
continuity in the sequence of surface brightness profile shapes as a
function of luminosity between the two sequences (e.g.,~Jerjen \&
Binggeli 1997; Graham \& Guzm{\'a}n 2003; Ferrarese \etal\ 2006).  In
our view the controversy has been definitively settled by the
comprehensive photometric study of Kormendy \etal\ (2008; see also
e.g.,~Geha \etal\ 2002), which demonstrates that elliptical and
spheroidal galaxies are cleanly separated in the fundamental plane
projections, even in the region of luminosity overlap between these
two classes of galaxies.  The implication of this distinction is of
critical importance in the context of our study; it suggests that
spheroidal galaxies do not share the violent merger history of
elliptical galaxies, but rather formed in a manner similar to disk
galaxies (e.g.,~Kormendy 1985; Bender \etal\ 1992).  In this
interpretation, spheroidals and pseudobulges presumably have more in
common than spheroidal and elliptical galaxies.  We are in a position
to evaluate (a) whether there is a population of galaxies without
classical bulges that nevertheless host central BHs and (b) how the
relationship between BH mass and ``bulge'' mass depends on the host
galaxy structure.  In this paper we follow Kormendy \etal\ (2008) and
explicitly use the term ``spheroidals'' to describe systems that in
the literature are often called ``dwarf ellipticals''.

Throughout we assume the following cosmological parameters to calculate
distances: $H_0 = 100~h = 71$~\kms~Mpc$^{-1}$, $\Omega_{\rm m} = 0.27$,
and $\Omega_{\Lambda} = 0.75$ (Spergel \etal\ 2003).

\section{The Sample}

Our intermediate-mass BHs are selected from the SDSS sample of AGNs
with broad \halpha\ ($z<0.352$), described in detail in Greene \& Ho
(2004).  Throughout we will refer to the targets using the
identification numbers assigned in that paper.  The 19 AGNs in this
sample have been followed up extensively in the radio (Greene \etal\
2006), X-ray (Greene \& Ho 2007a; Miniutti \etal\ 2008),
and mid-infrared with \spitzer.  Furthermore, most have stellar
velocity dispersions measurements taken with ESI on Keck (Barth \etal\
2005).  The sample is selected to have BH masses \mbh$< 10^6$~\msun\
and to have minimal galaxy continuum within the SDSS aperture.  One of
the primary goals of our original search was to investigate the host
galaxy properties of a BH mass-selected sample of low-mass BHs.  We
pursue that goal here.  

It is worth noting, however, that there are implicit constraints on
the host galaxy light set by the selection process, which are
discussed in detail by Greene \& Ho (2007b).  To be spectroscopically
targeted by the SDSS the galaxy must be relatively bright, while to be
spectroscopically identified as a broad-line AGN based on the \halpha\
profile the host galaxy must not swamp the weak emission lines.  When
we model these constraints, we find a range of possible host galaxy
luminosities that is very similar to that of the observed
distribution.  Although our search was designed to be free of host
galaxy bias, in fact there is a strong implicit host galaxy luminosity
bias built into our selection process, which must be considered when
interpreting our results.  Nevertheless, despite this major caveat, it
is still of considerable interest to examine the host galaxy
properties of this unique sample.

Since the sample selection is based on BH mass, we must also discuss
our BH mass measurement technique.  Of course, dynamical techniques
that use the motions of stars or gas in the vicinity of the BH
(e.g.,~Gebhardt \etal\ 2000a) provide the most accurate BH mass
measurements.  However, as discussed above, the spatial resolution
requirements are beyond current capabilities, and thus we are forced
to rely on less direct mass indicators.  In this case the dense gas
orbiting the BH (the broad-line region or BLR) is assumed to be
virialized and is used as a dynamical tracer.  The BLR velocity
dispersion is derived from the width of a broad line (in our case
\halpha; Greene \& Ho 2005b).  BLR sizes have been measured for a
handful of nearby AGNs by measuring the lag between variability in the
AGN continuum and corresponding variability in the line emission
(reverberation mapping; Blandford \& Mckee 1982; Peterson \etal\
2004).  With measured BLR radii, it is possible to calibrate an
empirical relation between the AGN luminosity and the BLR size
(e.g.,~Kaspi \etal\ 2005; Bentz \etal\ 2006), and from this relation
to estimate BH masses from single-epoch spectroscopy.  We will refer
to masses measured in this manner as ``virial'' BH masses.

A host of potentially very serious systematic uncertainties are
inherent in this technique (e.g.,~Krolik 2001; Collin \etal\ 2006),
particularly since the AGNs in our sample are considerably less
luminous, and harbor less massive BHs, than the bulk of the
reverberation mapping sample.  Nevertheless, in the cases for which
they are available, \sigmastar\ measurements show reassuring agreement
with both reverberation-mapping masses (Gebhardt \etal\ 2000b;
Ferrarese \etal\ 2001; Onken \etal\ 2004; Nelson \etal\ 2004) and
the virial masses of significantly larger samples of AGNs (Greene
\& Ho 2006b; Shen \etal\ 2008a).

Finally, the technical measurement details can seriously and
systematically impact the derived BH masses (e.g.,~Collin \etal\ 2006;
Shen \etal\ 2008b).  Note, for instance, that the BH masses presented
herein differ from those in the original Greene \& Ho (2004) paper
(and in some cases are greater than the original mass limit of
$10^6$~\msun), both because of differences in line width measurements
between the SDSS and ESI spectra (Barth \etal\ 2005) and because of
revision to the relation between BLR radius and AGN luminosity (Bentz
\etal\ 2006).  In this paper, we will use the \halpha\ calibration of
Greene \& Ho (2005b), and the most recent calibration of the
radius-luminosity relation presented by Bentz \etal\ (2006).  We
explain our measurement procedure in detail in the Appendix of Greene
\& Ho (2007c), and the interested reader is directed there for more
information.

\section{Observations and Methods}

We were awarded 19 orbits in Cycle 14 to observe the 19 Greene \& Ho
(2004) targets with \hst/ACS.  The observation dates and durations are
shown for each target in Table 1; GH03 is excluded due to a guide-star
acquisition failure and will not be discussed further.  Each orbit was
equally divided between F814W (broad $I$ band) and F435W ($B$ band),
but due to saturation concerns, the F814W observation was split
between a short (30 s) and long ($\sim 900$ s) exposure.  Although the
\hst\ and Johnson filters are not identical, for convenience we refer
to F435W as $B$ and F814W as $I$ throughout the 
manuscript\footnote{ We often compare our results to fiducial colors
of galaxies as a function of Hubble type using the synthetic
photometry of Fukugita \etal\ (1995).  Although that photometry was
performed in the Johnson $B$ and $I$ filters, the differences between
these and the \hst\ filters are $<0.05$ mag for all spectral types (as
we verified using {\it synphot} in IRAF).  Since intrinsic variations in color
due to reddening and stellar populations are always larger than this,
we do not worry further about the filter differences.}. Both for
cosmic ray removal and improvement of the final image sampling, the
images were taken in two dither steps, using a standard ACS two-step
dither pattern.  The $I$-band images are the main focus of this paper;
because of the high throughput and red color of this wide filter, it
is optimal for probing host galaxy structure.  On the other hand, the
$B$-band images provide important constraints on the presence and
morphology of ongoing star formation.  Furthermore, because the AGNs
are blue, we are able to derive our most robust measurements of the
AGN luminosity from these images, which leads to a significant
improvement in our estimate of the BH mass.

Data reduction (bias subtraction and flat-fielding) was performed by
the ACS CALACS software pipeline (Pavlovsky \etal\ 2005).  The
dithered images are combined and cleaned of cosmic rays using the
MULTIDRIZZLE software (Koekemoer \etal\ 2002), which linearly
reconstructs the image, accounting for geometric distortion (Fruchter
\& Hook 2002).  We make only minor changes to the default parameters.
One is to perform our own cosmic ray rejection using software written
by P.~Martini.  The other is to keep the native orientation of the
images (as opposed to specifying North up in the final images).  This
latter step is necessary to ensure that the stars we use as models of
the point-spread function (PSF) are oriented properly across all
images.  No sky subtraction is performed at this stage.  The resultant
images have a resolution of FWHM$=$0\farcs11, as measured from bright
stars in the $I$-band images, and a corresponding typical physical
resolution of $\sim 170$~pc at the median $z \approx 0.08$.  We reach
a typical 10~$\sigma$ limiting surface brightness of 23.7 mag
arcsec$^{-2}$ in the $I$ band.  All magnitudes are reported in the Vega
system.

%%%%%%%%%%%%%%%%%%%%%%%%%%%%%%%%%%%%%%%%%%%%%%%%%%%%%%%%%%%%%%%%%%%%
%%BoundingBox: 
\begin{figure*}
\vbox{ 
\hskip 0.4in
\psfig{file=tableobsv5.epsi,width=0.9\textwidth,keepaspectratio=true,angle=0}
}
\end{figure*}
\vskip 4mm
%%%%%%%%%%%%%%%%%%%%%%%%%%%%%%%%%%%%%%%%%%%%%%%%%%%%%%%%%%%%%%%%%%%%%
%\noindent

\subsection{The Point-Spread Function}

One of the major benefits of \hst\ is, of course, the very stable PSF.
However, even in the absence of atmosphere, the \hst/ACS PSF is known
to vary as a function of both detector position and time
(e.g.,~Sirianni \etal\ 2005; Rhodes \etal\ 2007).  Note that for our
application the PSF shape is of particular importance due to the
presence of a strong nuclear point source.  For the $I$ filter, we
have reconstructed PSF models from three bright stars in the fields of
GH02, GH16, and GH17, respectively.  The stars are significantly
brighter than any of the AGNs; they are saturated in the long
exposure, but luckily we have a measurement of the core from the 30 s
image.  These PSF stars were observed at different times (between
Oct. and Nov. of 2005) and are at different positions ($\sim$~300,
800, and 600 pixels from the typical object position, respectively).
We also use the TinyTim software (Krist 1995), which models both the
spatial and spectral variations in the PSF, and accounts for effects
such as charge diffusion that tend to broaden the PSF.  While it would
be ideal to have a PSF star observation for each object, we can at
least quantify the systematic uncertainty arising from PSF mismatch as
the differences between fits using each PSF star.  For the F435W
filter, we rely exclusively on TinyTim models and an IDL wrapper
(Rhodes \etal\ 2006, 2007) that further generates PSF models with
focus values of $\pm 5 \micron$ around the nominal focus, to use the
terminology of Rhodes \etal\

\subsection{Non-parametric Magnitudes}

Perhaps the most basic measurement we might make is that of the total
host galaxy and AGN luminosity in each band.  For this purpose we need
not assume any model for the host galaxy light; we need only a model
of the PSF.  Prior to performing detailed fits to the two-dimensional
image profiles, we perform non-parametric measurements of the AGN and
host galaxy magnitudes.

The basic premise in deriving non-parametric magnitudes is that within
a small enough aperture at the center of the galaxy, the AGN
completely dominates the luminosity.  Thus one simply scales the model
PSF to match the AGN flux within this small aperture, and subtracts
the scaled PSF magnitude.  To the extent that the PSF model is an
accurate representation of the light distribution of the AGN, aperture
effects are automatically accounted for.  This method fails
systematically, however, when the AGN does not dominate the light
within the chosen aperture.  It is desirable to choose the smallest
possible aperture to mitigate contamination from the host galaxy, in
principle as small as a single pixel.  Unfortunately, using a single
pixel makes the procedure very sensitive to centering errors.  For
that reason, we follow Jahnke \etal\ (2004) and measure the unresolved
nuclear magnitude within an aperture of radius 2 pixels.  We perform
the photometry using the package {\it phot} within IRAF and apertures
of 1--200 pixels in radius.  The sky is subtracted from each aperture
as determined from an aperture of radius 220 pixels and a width of 20
pixels.  Again, of primary concern is the systematic overestimate of
the AGN luminosity that necessarily ensues due to our assumption that
it accounts for 100\% of the light within the inner pixels.  The
appropriateness of this assumption depends on both the contrast
between the AGN and the host galaxy, and the light distribution of the
host galaxy.  Centrally concentrated hosts are the most difficult to
uncover.

The outer apertures are incremented in five-pixel steps until the
magnitude is stable to within 0.05 mag.  These non-parametric
magnitudes, both for the AGN and the total galaxy luminosity are
presented in Table 2.  It is clear from this exercise that, as claimed
in Greene \& Ho (2004), the host galaxies are generally much brighter
than the AGNs, unlike luminous AGNs, whose host galaxies may account
for only $\sim 10\%$ of the total luminosity of the system.  Even
using the non-parametric PSF-scaled luminosities, which may be viewed
as an upper limit on the total nuclear strength, we find that on
average the AGN accounts for only 30\% of the total luminosity in the
$B$ band and $20\%$ of the total luminosity in the $I$ band.  Only in
GH15 does the nuclear luminosity exceed the galaxy luminosity.
Therefore, in the majority of cases we should be able to measure
robust galaxy structural parameters for these galaxies.  We compare
the results of the parametric and non-parametric photometry below, and
find reassuringly good agreement.

\subsection{GALFIT}

Non-parametric measurements provide a valuable benchmark for our
subsequent work since they are model-independent.  Still, we wish to
parametrize the structures of our host galaxies, for comparison with
inactive local galaxies.  Furthermore, since all fit parameters are
coupled, without a model for the inner galaxy profile it is not
possible to get a proper measurement of the AGN magnitude.  For this
reason, we proceed now to parametrized fitting.

Generally speaking, our goal is to decompose the observed intensity
distribution into a combination of components including a central
unresolved point source, and some combination of a possible disk,
bulge, and bar.  In general profile-fitting can be performed either
with azimuthally averaged profiles or full two-dimensional images.
Fitting the full two-dimensional profile can provide important
additional constraints (for instance in the presence of ellipticity
changes with radius), and in the presence of multiple complex
components (such as nuclei and bars) this flexibility is of particular
importance (e.g.,~Andredakis \etal\ 1995; Byun \& Freeman 1995,
Wadadekar \etal\ 1999).  Here we utilize the versatile GALFIT routine
to perform two-dimensional profile decompositions (Peng \etal\
2002). GALFIT has been used and tested extensively for a variety of
applications including AGN host-galaxy work (e.g.,~Ho \& Peng 2001;
Peng \etal\ 2006a, 2006b; Kim \etal\ 2007), and galaxy morphology work
at low (e.g.,~Peng 2002) and high (e.g.,~Bell \etal\ 2006) redshift.

Briefly, the current version of GALFIT models each galaxy component as
an azimuthally symmetric ellipsoid, with a number of possible radial
intensity profiles, including the generalized \sers\ (1968) model
\begin{equation}
I(r) = I_e~{\rm exp} \left[ -b_n \left(\frac{r}{r_e}\right)^{1/n}-1 \right],
\end{equation}
\noindent
where $r_e$ is the effective (half-light) radius, $I_e$ is the 
intensity at $r_e$, $n$ is the \sers\ index, and $b_n$ is chosen 
such that
\begin{equation}
\int_0^{\infty}I(r) 2 \pi r dr = 2 \int_0^{r_e} I(r) 2 \pi r dr.
\end{equation}
We adopt the analytic approximation for $b_n$ from MacArthur \etal\
(2003), as adapted from Ciotti \& Bertin (1999):  
\begin{equation}
b_n \approx 2n-\frac{1}{3}+\frac{4}{405n}+\frac{46}{25515n^2}+
\frac{131}{1148175n^3}-\frac{2194697}{30690717750n^4}.
\end{equation}
The \sers\ model is
particularly convenient, as it reduces to an exponential profile for
$n=1$ and a de Vaucouleurs (1948) profile for $n=4$.  Bars may be
modeled as ellipsoids with very low axial ratios.  We follow de Jong
(1996a; see also Freeman 1966) and model the intensity distribution in
the bar as a Gaussian ($n=0.5$).  In total for a given \sers\ component, the
model parameters include the two-dimensional component centroid, the
total magnitude, the \sers\ index, the effective radius, the position
angle, and the ellipticity (which are constants with radius for a given
component in this version of GALFIT).  Parameter estimation is
performed using Levenberg-Marquardt minimization of \chisq\ in pixel
space.  The sky is modeled as a constant pedestal offset.  The effects
of the telescope optics are accounted for by convolution with a
user-provided model of the instrumental PSF, and foreground objects or
detector gaps are explicitly masked with a user-defined bad-pixel
list.  In general, for the single-component models predominantly
discussed here, GALFIT is able to converge rapidly with only
moderately reasonable input parameters.

Frequently the \hst\ PSF is actually undersampled.  The problem is
most severe for WFPC2, but even ACS/WFC, with 0\farcs05 pixels, is
undersampled.  As a result, centering mismatch occurs at a level that
leads to significant parameter estimation errors.  M.~Kim \etal\ (in
preparation) use simulations to show that significant systematic
uncertainty can result from the impact of undersampling.  They
recommend broadening the image and PSF with a $\sigma=2$~pixel
Gaussian kernel, which they show largely mitigates the systematic
offsets in their simulations.  Therefore, we also perform the fitting
with a broadened PSF star, using the stars from the GH02 and GH16
observations, which tend to provide the best fits.

%%%%%%%%%%%%%%%%%%%%%%%%%%%%%%%%%%%%%%%%%%%%%%%%%%%%%%%%%%%%%%%%%%%%
%%BoundingBox: 
\begin{figure*}
\vbox{ 
%\vskip -0.1truein
\hskip -0.1in
\psfig{file=tablegalv6.epsi,width=0.5\textwidth,keepaspectratio=true,angle=-90}
}
\end{figure*}
%\vskip 4mm
%%%%%%%%%%%%%%%%%%%%%%%%%%%%%%%%%%%%%%%%%%%%%%%%%%%%%%%%%%%%%%%%%%%%%
%\noindent

%%%%%%%%%%%%%%%%%%%%%%%%%%%%%%%%%%%%%%%%%%%%%%%%%%%%%%%%%%%%%%%%%%%%
%%BoundingBox: 
\begin{figure*}
\vbox{ 
\vskip -0.1truein
\hskip 0.in
\psfig{file=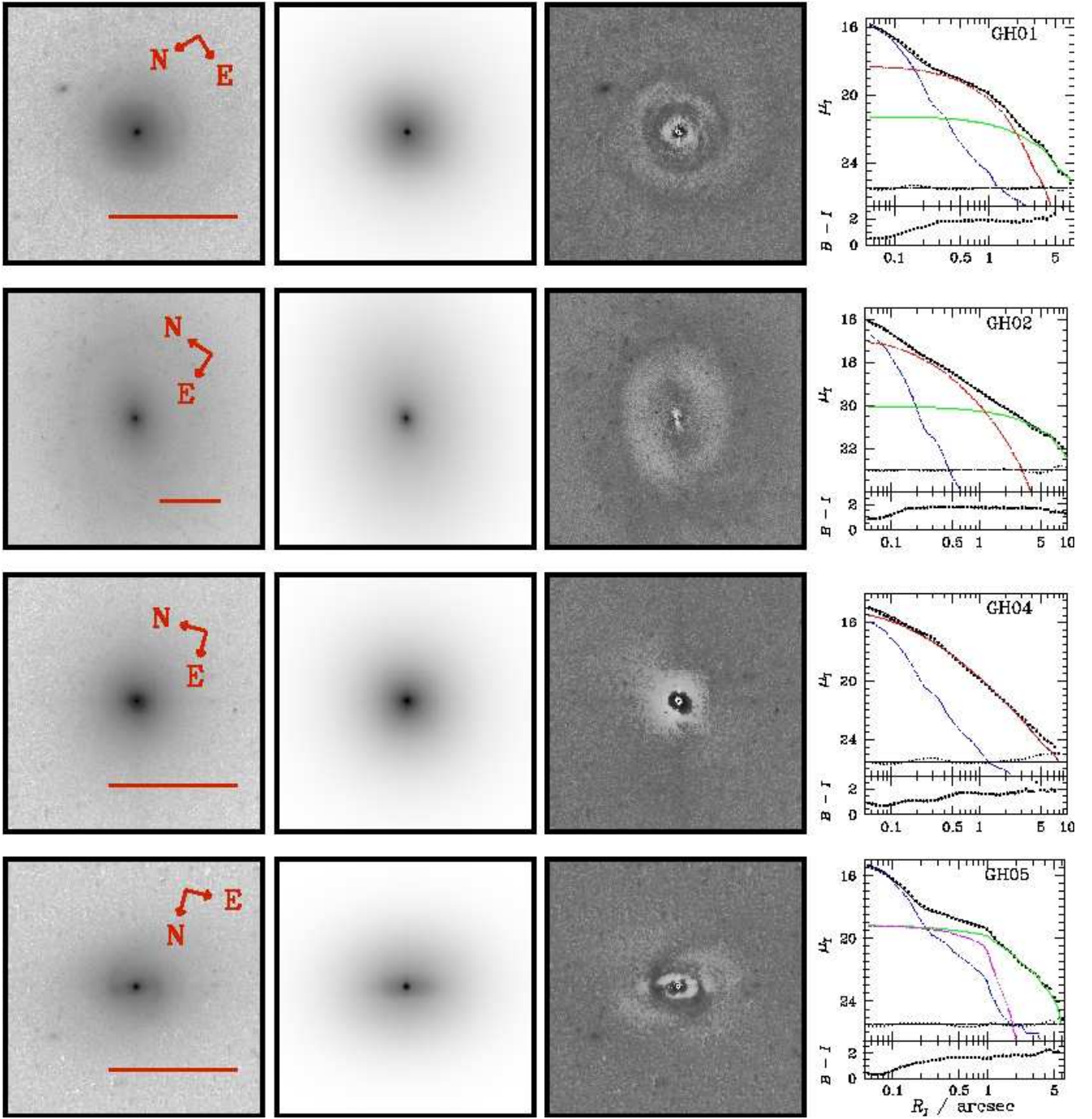,width=1.0\textwidth,keepaspectratio=true,angle=0}
}
\vskip -0mm
\figcaption[]{
\hst\ images presented in this paper.  From left to right we display
the $I$-band image, the preferred GALFIT model, the residuals, and an
azimuthally averaged profile created with the IRAF program {\it
ellipse}.  Scale bar represents 5\arcsec.  In the two-dimensional
residuals there is often a nearly vertical artifact caused by
diffraction spikes in the star used to model the AGN.  Plotted in the
one-dimensional residuals are the data ({\it filled circles}), the
total model ({\it solid line}), the AGN ({\it blue long-dash-dotted
line}), and, if present, a bulge ({\it red dotted line}), a disk ({\it
green dash-dot line}), and bar ({\it magenta dash-dot line}).  The fit
residuals are shown as small dots.  In the bottom panel we show the
$B-I$ color profile of the galaxy, including the AGN.
\label{fits}}
\end{figure*}
%\vskip 5mm
%%%%%%%%%%%%%%%%%%%%%%%%%%%%%%%%%%%%%%%%%%%%%%%%%%%%%%%%%%%%%%%%%%%%%
%%%%%%%%%%%%%%%%%%%%%%%%%%%%%%%%%%%%%%%%%%%%%%%%%%%%%%%%%%%%%%%%%%%%
%%BoundingBox: 
\begin{figure*}
\vbox{ 
\vskip -0.1truein
\hskip 0.in
\psfig{file=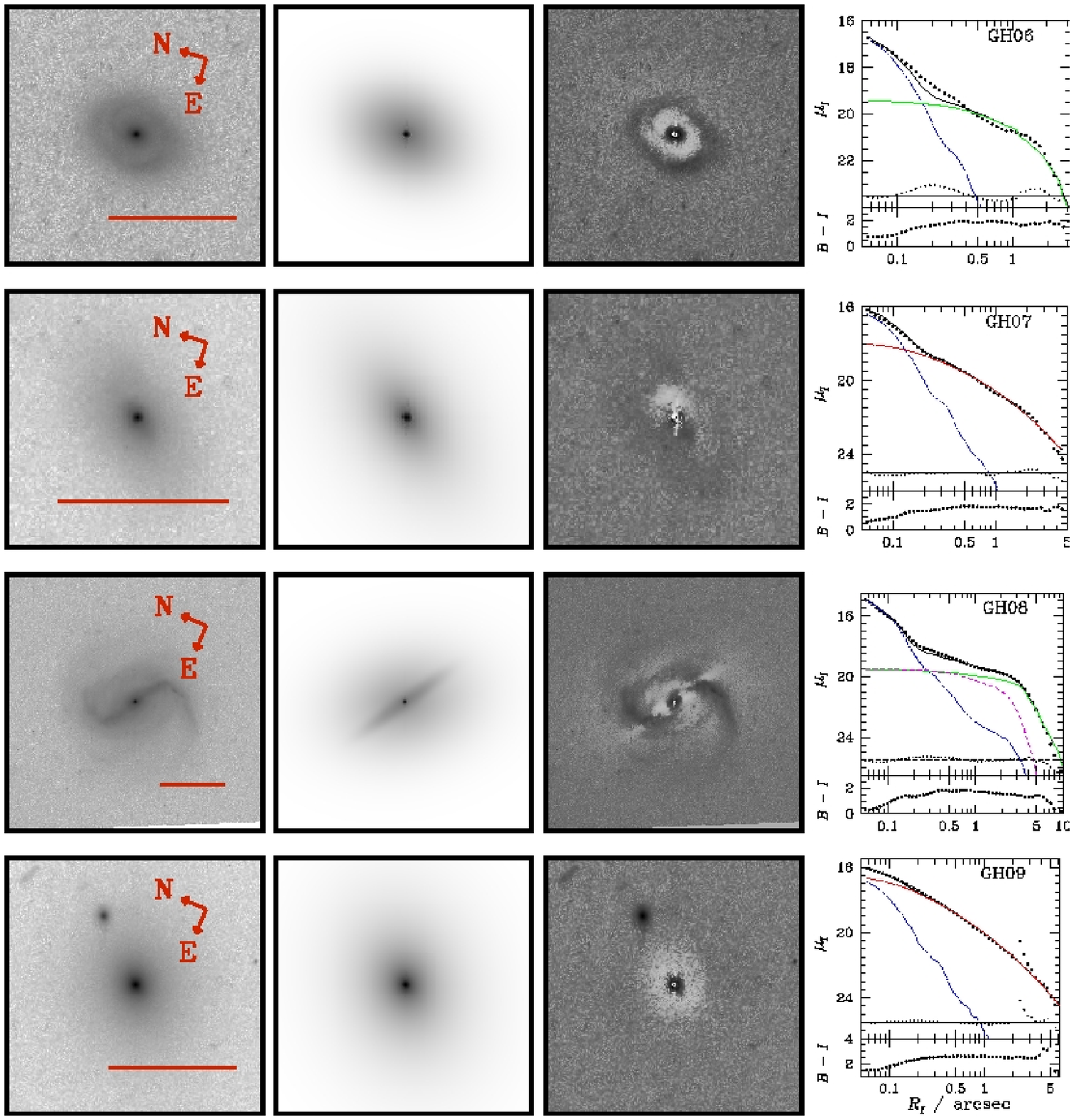,width=1.0\textwidth,keepaspectratio=true,angle=0}
}
Fig. 1. --- continued.
\end{figure*}
%%%%%%%%%%%%%%%%%%%%%%%%%%%%%%%%%%%%%%%%%%%%%%%%%%%%%%%%%%%%%%%%%%%%
%%%%%%%%%%%%%%%%%%%%%%%%%%%%%%%%%%%%%%%%%%%%%%%%%%%%%%%%%%%%%%%%%%%%
%%BoundingBox: 
\begin{figure*}
\vbox{ 
\vskip -0.1truein
\hskip 0.in
\psfig{file=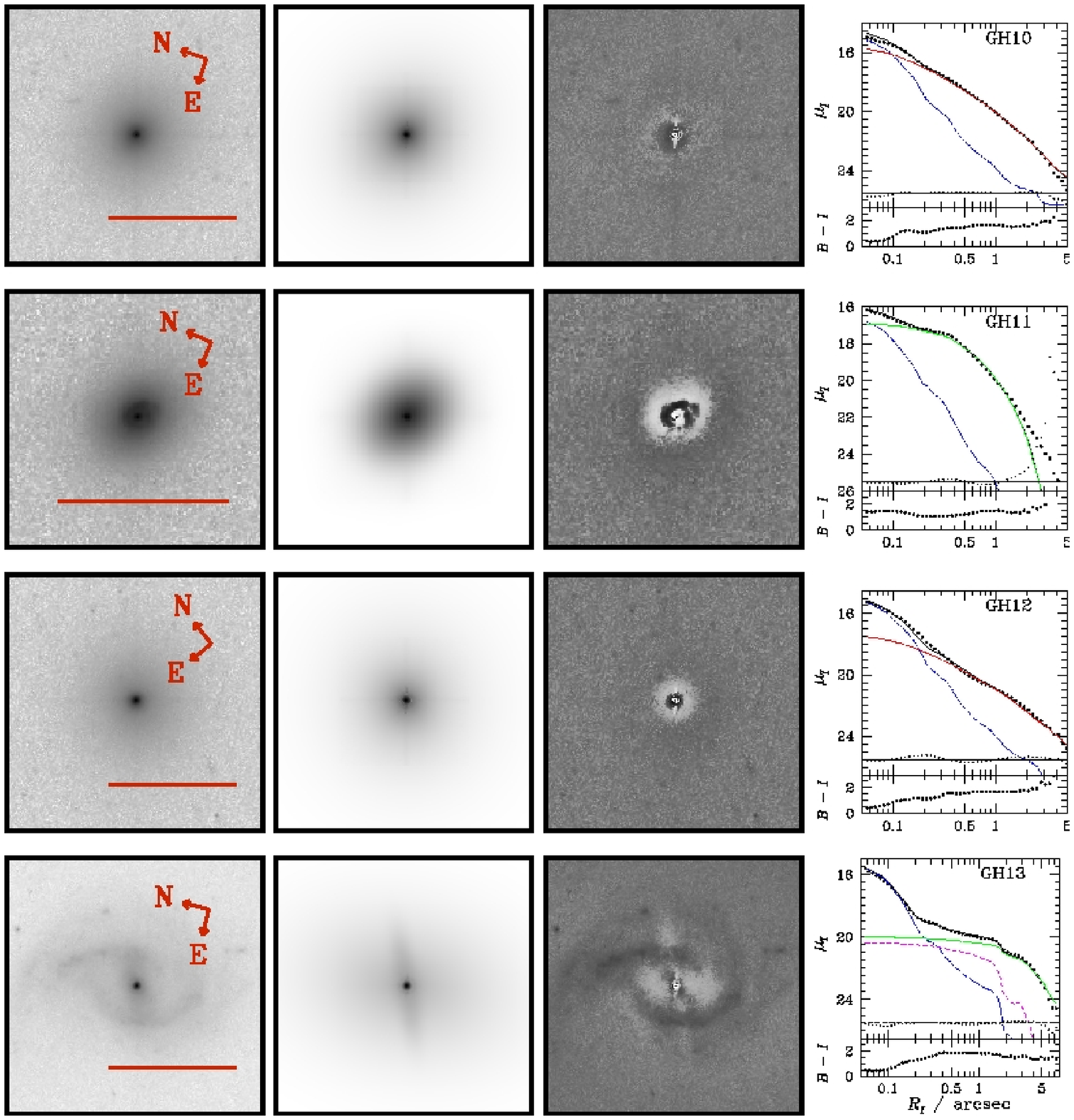,width=1.0\textwidth,keepaspectratio=true,angle=0}
}
Fig. 1. --- continued.
\end{figure*}
%%%%%%%%%%%%%%%%%%%%%%%%%%%%%%%%%%%%%%%%%%%%%%%%%%%%%%%%%%%%%%%%%%%%
%%%%%%%%%%%%%%%%%%%%%%%%%%%%%%%%%%%%%%%%%%%%%%%%%%%%%%%%%%%%%%%%%%%%
%%BoundingBox: 
\begin{figure*}
\vbox{ 
\vskip -0.1truein
\hskip 0.in
\psfig{file=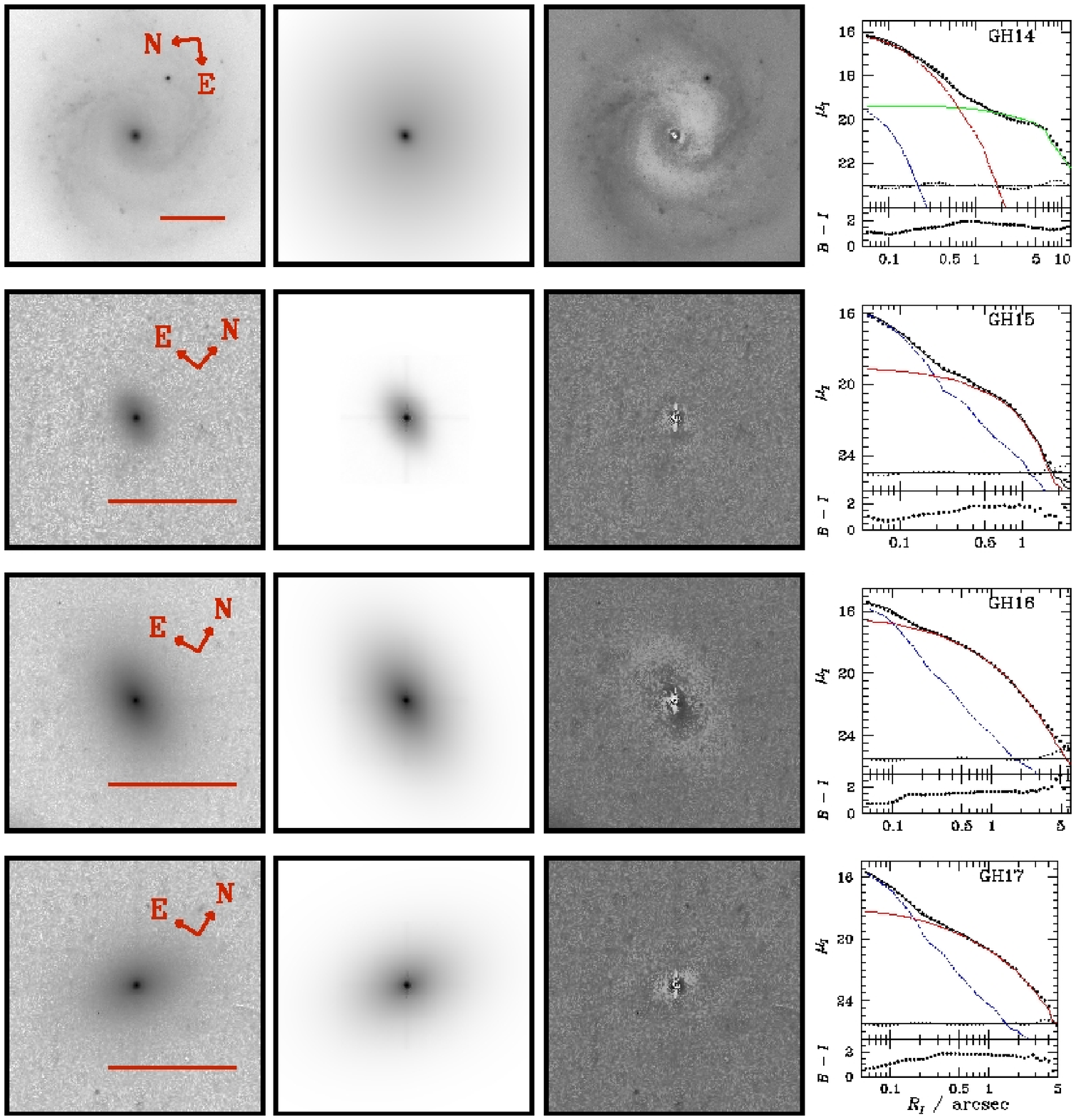,width=1.0\textwidth,keepaspectratio=true,angle=0}
}
Fig. 1. --- continued.
\end{figure*}
%%%%%%%%%%%%%%%%%%%%%%%%%%%%%%%%%%%%%%%%%%%%%%%%%%%%%%%%%%%%%%%%%%%%
%%%%%%%%%%%%%%%%%%%%%%%%%%%%%%%%%%%%%%%%%%%%%%%%%%%%%%%%%%%%%%%%%%%%
%%BoundingBox: 
\begin{figure*}
\vbox{ 
\vskip -0.1truein
\hskip 0.in
\psfig{file=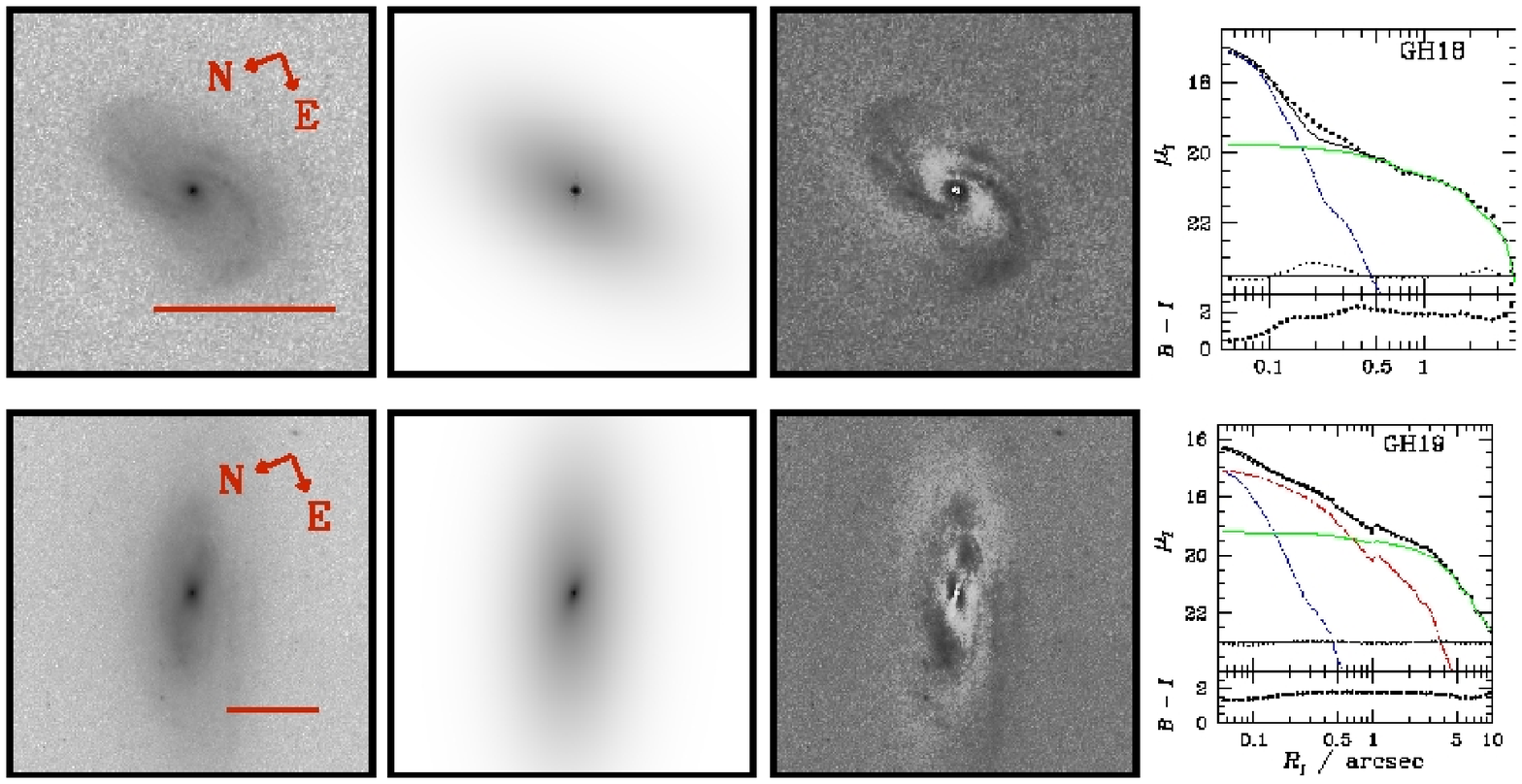,width=1.0\textwidth,keepaspectratio=true,angle=0}
}
Fig. 1. --- continued.
\end{figure*}
%%%%%%%%%%%%%%%%%%%%%%%%%%%%%%%%%%%%%%%%%%%%%%%%%%%%%%%%%%%%%%%%%%%%

In what follows, we discuss the structural properties of the host
galaxies as derived from GALFIT models of the $I$-band images, since the
host galaxy contrast is maximized in this band.  The AGN is modeled as
a scaled version of the PSF, while the host galaxies are modeled as
\sers\ components.  We have performed fits to the $B$-band images as
well, in which all of the structural measurements are fixed to the
$I$-band fits, but the magnitudes are allowed to vary.  The $B$-band
fits, while shallow, provide our most robust measurement of the AGN
luminosity.  We do not have a strong handle on the intrinsic reddening
nor on the true galaxy color, and so we do not assume a particular
value for the reddening.

\section{Results}

Even in the absence of a bright nuclear point source, profile
decomposition is a difficult endeavor, and without sufficient depth
and angular resolution parameter coupling renders the extraction of
multiple components unreliable (e.g.,~Kormendy \& Djorgovski 1989;
MacArthur \etal\ 2003).  The active nucleus only accentuates these
difficulties.  Therefore, while GALFIT in principle can solve for any
number of completely general \sers\ components, we are forced to
consider a much more restricted set of models.  First, we divide the
sample into those systems with a clear disk component\footnote{There
are two galaxies for which this distinction was not obvious.  GH11 is
morphologically disturbed but is best fit as a compact disk (see
Appendix), while GH15 is simply very faint.  There is really no
preference between the $n=1$ and $n=2$ models for GH15, but the former
has a marginally smaller \chisq.}  (GH01, GH02, GH05, GH06, GH08,
GH13, GH14, GH15, GH18, GH19), and those that are simply ``blobs'' (we
agnostically refer to compact, smooth galaxies as blobs, since until
we examine their structures in detail we do not know whether they are
elliptical or spheroidal galaxies; see Fig. 1).  We fit the profile
with a minimum number of components, and then place limits on the
existence of a second component (disk or bulge for the blob or disk
galaxies, respectively).  We make the further assumption that the
underlying galaxies obey structural scaling relations akin to inactive
systems, and use these scalings to limit parameter space when deriving
limits.  Since the details are different for the blob and disk
systems, we discuss each subsample separately below.

\subsection{Blobs}

Particularly in the case of steep \sers\ profiles, a nuclear point
source complicates our ability to recover the true index $n$.  For
this reason, rather than allowing the \sers\ index to float, we simply
run a grid of models with $n=1,2,3,4$, and then ask which does the
best job at fitting the data.  The \chisq\ values returned by GALFIT,
while they have some relative value, do not have any absolute
discriminatory power, because we are dominated by systematic
uncertainties (\S 4.3).  To assist in determining the best-fit \sers\
index, we also compute a \chisq\ from radial profiles of both the
best-fit model and the data constructed using the program {\it
ellipse} within IRAF, which is based on the techniques described in
Jedrzejewski (1987).  We report best-fit values averaged over all PSF
stars, and we include in the error budget the range of luminosities
and sizes that ensues from a range in \sers\ index of $n \pm 1$.
While we might expect these low-luminosity galaxies to have relatively
low \sers\ indices based on the correlation between galaxy luminosity
and \sers\ index (e.g.,~Schombert 1986; Caon \etal\ 1993; Graham
\etal\ 1996; MacArthur \etal\ 2003), many of the blobs are best fit by
classic de Vaucouleurs ($n=4$) profiles.  The same is true of POX 52 
(Thornton \etal\ 2008).

It is not completely clear that single-component models are a good fit
for all of the blobs. Particularly in the case of GH04 and GH12 there
appears to be a compact ($\sim 0\farcs15-0\farcs3$ or hundreds of pc
at the distance of these sources) residual.  There are a variety of
possible explanations for this component.  The most natural is PSF
mismatch, since there is real structure in the PSF on this scale.
However, the compact residuals persist in these two systems when we
perform the fits on the smoothed images, suggesting that PSF mismatch
is not the culprit.  If this is indeed a physical component, we note
that it is significantly more extended than the nuclear star clusters
that appear to be a common component of the nuclei of galaxies of both
late (e.g.,~B{\"o}ker \etal\ 2002) and early type (e.g.,~Carollo
\etal\ 2002; C{\^o}t{\'e} \etal\ 2006).  Nuclear rings of star
formation can be this size (e.g.,~Buta \& Combes 1996), and the colors
of GH04 and GH12 are bluer than typical elliptical galaxies.
Interestingly, Thornton \etal\ (2008) also see evidence for a
similar component in the $B$-band image of POX 52. Given the potential
complications of PSF mismatch, we are wary of overinterpreting the
physical significance of the residuals.

While the blob images do not present any compelling visual evidence
for an additional exponential disk, we wish to place limits on the
presence of such a component.  There is an infinite family of
possible low surface brightness disks, but normal disks do not span
this entire space.  Rather, there is an empirical correlation between
disk scale length and bulge size.  This correlation has been seen in a
various studies (e.g.,~Courteau \etal\ 1996; de Jong \etal\ 1996b;
MacArthur \etal\ 2003), at a variety of wavelengths, and over a broad
range in Hubble type.  By constraining the putative disk component to
obey these scalings we can place limits on the existence of typical
disk components in the blob galaxies.

Since our galaxies are relatively faint, we use the relation from
MacArthur \etal\ that was calibrated for late-type spiral galaxies;
$\langle r_e/h \rangle=0.22 \pm 0.09$, where $r_e$ is the effective
radius of the bulge and $h$ is the scale length of the disk (and the
effective radius of the disk, $r_d=1.678h$).  In addition to the
scatter in the relation, there is significant uncertainty in the real
blob parameters.  Folding in all of these uncertainties yields a
maximum and minimum disk size.  At the same time, we need to assign a
central surface brightness to each disk; we use the relation between
bulge and disk surface brightness shown in Fig. 19 of de Jong (1996b)
with a magnitude of scatter in each direction.  In converting between
$B$ and $I$, we use a color $B-I=1.67$ mag for Scd galaxies (Fukugita
\etal\ 1995).  Each size and central luminosity corresponds to a
well-defined radial profile, assuming an exponential disk and an
inclination of 45\degr.  To test the detectability of this disk, we
simply calculate the radius at which the surface brightness of the
disk and blob are equivalent.  As long as that radius is within the
limiting radius at which we detect the blob (as determined from the
radial profiles in Fig. 1), we would detect the disk if it were
present.  By this criterion, all the simulated disks are detectable
and therefore we can state rather strongly that the blob galaxies do
not have a ``normal'' disk component.  Note that it is much harder to
limit the presence of a face-on disk.  In that case we still might
expect to see spiral structure in the fit residuals, which we do not.

\subsection{Disk Galaxies}

The basic approach is similar for the disk galaxies.  In this case
there is less degeneracy between the nucleus and the disk, which is
not centrally concentrated.  The best-fit disk is determined from
GALFIT with the \sers\ index fixed to 1.  As a first pass, we fit only
a disk component.  If there is a visible bar, as in the case of GH05,
GH08, and GH13, we then take the best-fit disk and add a bar with a
\sers\ index fixed to 0.5, which is equivalent to a Gaussian (e.g.,~de
Jong 1996a). In some other cases (GH01, GH02, GH14, GH19), there is a
clearly resolved bulge component, whose addition yields a dramatic
improvement in \chisq.  In other cases (GH06, GH18), although there
are hints of a bulge component in the residuals, the component is only
marginally resolved and we do not achieve stable fits with GALFIT.
For these cases, and those whose nuclei are dominated by the bar, we
place limits on the bulge component.  We note that in many cases the
disk galaxies in particular show substantial structure in the residual
images, typically corresponding to spiral arms, which we do not fit
(in some cases, such as GH08, it is also clear that our bar model is
not perfect).

For those systems without a clearly resolved bulge, our general
approach is similar to the case of the blobs.  Again, our strategy is
to limit the presence of a typical bulge, i.e.~one that obeys the
scaling relations seen in inactive galaxies.  We use the same
correlation between disk scale length and bulge effective radius from
MacArthur \etal\ (2003), and we further insist that the bulge lies on
the fundamental plane.  There are three bulge models that span the
range of sizes allowed by the measured disk-bulge size relation, and
from these sizes we assign a luminosity using the fundamental plane
projections of Bernardi \etal\ (2003), assuming a color transformation
of $i\arcmin - I = 0.63$ mag.  We repeat the fitting process including
a bulge with the sizes and luminosities fixed to the values thus
determined, and the spatial position, axial ratio, and position angle
fixed to that of the disk.  For increased sensitivity to changes in
the model on the scale of the bulge, we calculate two \chisq\
statistics from the radially averaged profiles.  The first uses all of
the data, while the second is limited to radii ranging from the
effective radius of the putative bulge to the measured effective
radius of the disk.  This is designed to maximize our sensitivity to
the regime in which light from the bulge would dominate, if present.
In most cases the \chisqr\ is close to unity for the total radial
profile, and in all cases the extra component makes a large impact on
the fits (sometimes improving it).  We adopt as the bulge upper limit
the largest (most luminous) bulge that cannot be ruled out from the
fitting.  The most important outcome of this exercise is that we are
unable to rule out the presence of a reasonably sized bulge-like
component in any of the galaxies.

We now have structural measurements (size, luminosity, and surface
brightness) or upper limits for the bulge-like components of all the
galaxies in our sample, and \sigmastar\ measurements for most of them.
In principle we can also derive color gradient information, which may
significantly impact our inferences about galaxy structure.  For
instance, a nuclear starburst would make the galaxy appear more
centrally concentrated than it truly is.  Unfortunately, the inner 200
pc of our images are dominated by the AGN.  The combination of
uncertainties in PSF shape (particularly in the blue) and centroiding
errors make it prohibitively difficult to extract believable color
information in the inner regions.  Before turning to an investigation
of the fundamental plane properties of these systems, we now briefly
investigate the most pernicious sources of systematic uncertainty and
their magnitudes.

%%%%%%%%%%%%%%%%%%%%%%%%%%%%%%%%%%%%%%%%%%%%%%%%%%%%%%%%%%%%%%%%%%%%
%%BoundingBox: 
\begin{figure*}
\vbox{ 
\vskip -0.1truein
\hskip 0.0in
\psfig{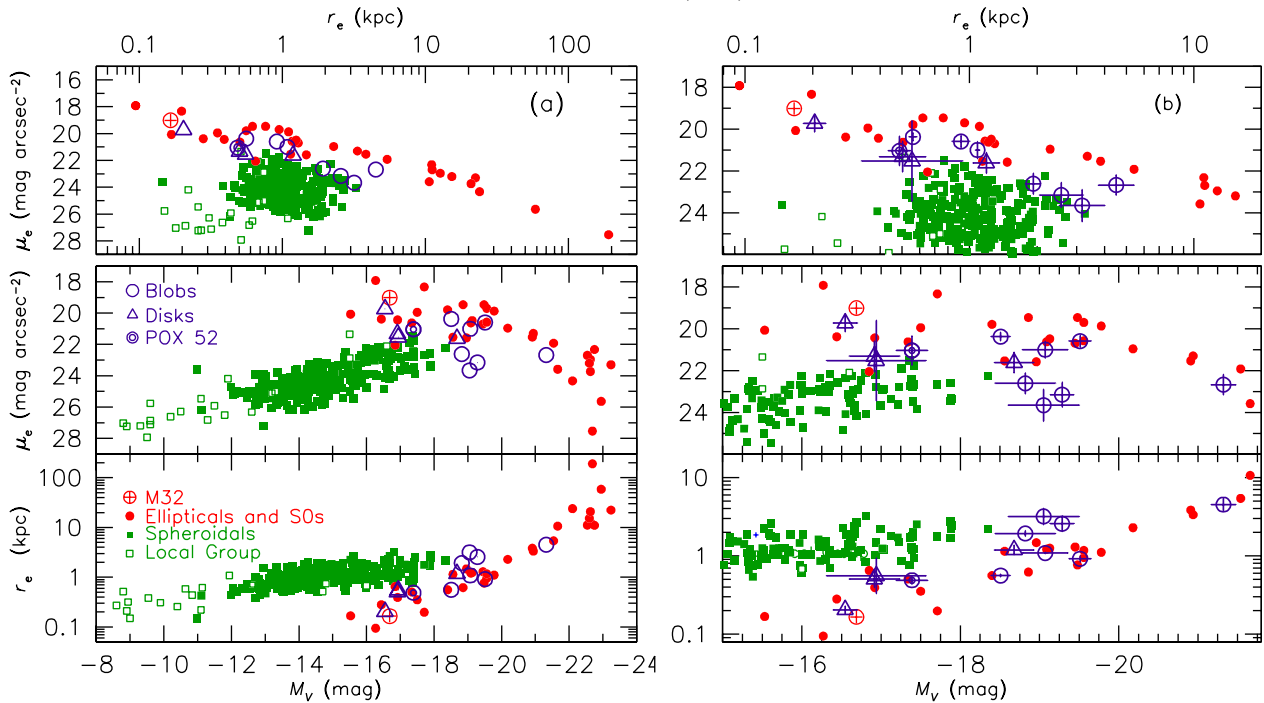}
}
\vskip -0mm
\figcaption[]{
({\it a}): The fundamental plane of hot stellar systems from the
recent study of Kormendy \etal\ (2008), predominantly for Virgo
cluster galaxies.  We show elliptical and S0 galaxies
({\it filled red circles}), Virgo spheroidal galaxies ({\it filled
green squares}; from Kormendy \etal\ 2008, supplemented by Ferrarese
\etal\ 2006 and Gavazzi \etal\ 2005) and Local Group dwarf spheroidal
galaxies ({\it open green squares}; Matteo 1998; McConnachie \& Irwin
2006).  Our systems are shown as open blue symbols, with the blobs as
circles and the disks with detected bulges as triangles.  POX 52 comes
from the recent \hst\ measurements of Thornton \etal\ (2008).  
Bulge upper limits are not included.
({\it b}): The same figure enlarged to highlight the
region where our objects are found.
\label{fpa}}
\end{figure*}
\vskip 5mm
%%%%%%%%%%%%%%%%%%%%%%%%%%%%%%%%%%%%%%%%%%%%%%%%%%%%%%%%%%%%%%%%%%%%%

\subsection{Uncertainties}

In this section we attempt to isolate the model assumptions and
parameters that potentially contribute most significantly to the error
budget for each fit.  The first is small errors in the sky estimation
that can lead to significant errors in magnitude and size
measurements.  To quantify the uncertainties due to sky level, we
simply rerun each GALFIT model with the sky level fixed at $\pm
1~\sigma$ of the best-fit value, where $\sigma$ is determined from
four well-separated regions of 200 square pixels each, and is
typically at the 1\% level but never exceeds 2\%.  The resulting
dispersion in parameter values is added in quadrature to the total
error budget of each parameter.  Typically the sky-related
uncertainties are at the 10\% level in radius, although they are
closer to 30\% for the bulges in disk galaxies, and $<0.1$ mag in
luminosity.

More difficult to quantify are uncertainties in the PSF model.  As
discussed above, while the \hst\ PSF is very stable, nevertheless it
displays both temporal and spatial variations (e.g.,~Rhodes \etal\
2007), and errors in the assumed PSF will translate into errors in the
inferred galaxy structural parameters.  Our approach is to use
multiple PSF models, derived both from bright stars in our fields and
from TinyTim, with the hope that the differences in the fits represent
the uncertainties incurred by imperfections in our model.  However, if
there is something systematically incorrect in all of our PSF models,
then we can underestimate the true errors.  One way to more robustly
quantify the impact of PSF variations is to perform simulations in
which the PSF model is changed in controlled ways.  M.~Kim et al. (in
preparation) have run a suite of simulations investigating the impact
of spatial, temporal, and color variation on the resulting fits.  For
simulations in which the host luminosity is $\sim 10$ times larger
than the nuclear luminosity, which is the case for the red images, PSF
mismatch does not impact the host galaxy fits substantially.  Indeed,
we find the variation in structural parameters from different PSF
models to be small ($<10\%$, with the exception of the bar
parameters).  At the same time, the spread in $I$-band AGN magnitudes
from different PSF stars can be as large as two magnitudes.

Although it is rarely discussed, Kim \etal\ point out that PSF
undersampling causes errors in centering that in turn lead to
systematic fitting errors.  While it is somewhat counterintuitive,
broadening both the PSF model and the image slightly can mitigate the
impact of undersampling. Therefore, we have repeated the fits on
images smoothed with a Gaussian kernel with $\sigma=2$ pixels in both
$x$ and $y$.  As expected, we find minimal changes to the host galaxy
properties, and this error contributes least to the error budget of
the structural parameters, but we do see typical variations in the PSF
magnitudes of $\sim 0.2$ mag.  We add this uncertainty in quadrature
to the errors of both the PSF and host galaxy models.

Finally, the hardest systematic uncertainties to quantify are those
due to our model assumptions, since in detail galaxies need not be
well-described by a single \sers\ model.  Even in the absence of a
nuclear point source, profile decomposition is a tricky game rife with
degeneracy (see, e.g.,~discussion in MacArthur \etal\ 2003).  We have
tried to be conservative in our error estimation by including the
range of measured properties from a wide range of \sers\ indices.
However, it is clear that in at least a few cases our models are not
completely adequate.  In addition to the clear nuclear residuals in
GH04 and GH12, GH11 also has well-resolved ring-like residuals.  It is
difficult to quantify the systematic errors associated with incorrect
models.  One thing we can say is that the total luminosities are
robust, based on comparisons with the non-parametric measurements.
The galaxy magnitudes show good agreement between the two methods,
with a median magnitude difference of $0.1 \pm 0.2$ mag in $I$, and a
median color difference of $-0.01 \pm 0.3$ mag.  Even the AGN
magnitudes agree well.  In $B$, the median magnitude difference is
$0.04 \pm 0.4$ mag.  The agreement is not as good in the $I$ band,
where the galaxy often constitutes a significant fraction of the total
light within the inner two pixels; the median non-parametric AGN
magnitude is $1 \pm 0.4$ mag brighter than the parametric fit.  This
is why we use the $B$-band fits to measure the AGN magnitudes.

\section{Host Galaxy Scaling Relations}

\subsection{The Fundamental Plane}

%%%%%%%%%%%%%%%%%%%%%%%%%%%%%%%%%%%%%%%%%%%%%%%%%%%%%%%%%%%%%%%%%%%%
%%BoundingBox: 
\begin{figure*}
\vbox{ 
\vskip -0.1truein
\hskip 0.in
\psfig{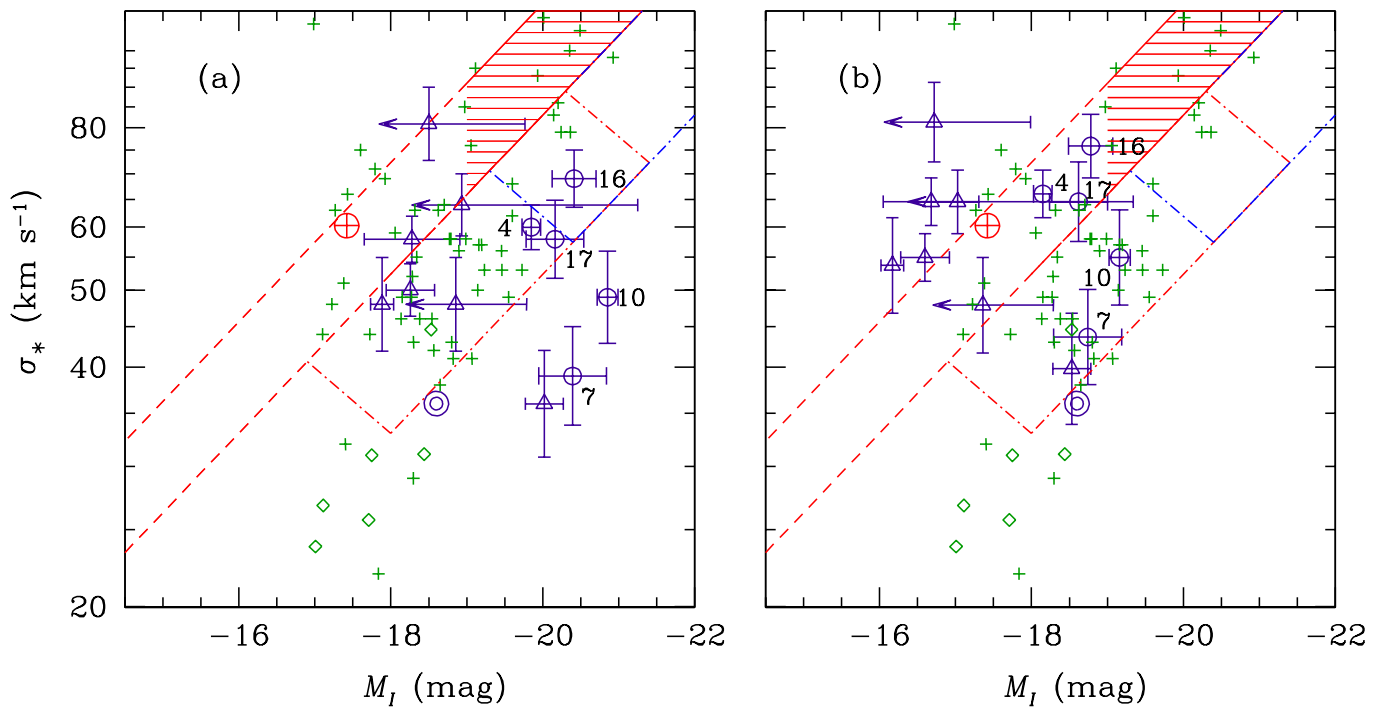}
}
\vskip -0mm
\figcaption[]{
({\it a}): 
Faber-Jackson projection of the fundamental plane.  As above, our
galaxies are shown as open blue symbols, with circles for blobs and
triangles for the disk galaxies with detected bulge components.  The
inactive relation for elliptical galaxies and classical bulges from
Bernardi \etal\ (2003; assuming $i-I=$0.63 mag) is shown as the red
shaded region, while the extrapolated region is shown as dashed lines.
Two samples of spheroidal galaxies are shown for comparison, Geha
\etal\ (2002; {\it green diamonds}, assuming $V-I=1.34$ mag) and
Matkovi{\'c} \& Guzm{\'a}n (2005; {\it green crosses}, assuming
$B-I=1.8$ mag).  From the data compilation in Kormendy \&
Kennicutt (2004), it appears that the relation may not be linear at the
lowest luminosities, but rather depart toward lower \sigmastar\ at a
given luminosity, as shown schematically by the red box in dash-dot.
Finally, from the same compilation we indicate the region occupied by
pseudobulges with a blue rectangle.  ({\it b}): Same as ({\it a}),
except that we have corrected the velocity dispersions for the
potential influence of either a nuclear star cluster or a face-on disk
component.  Furthermore, we have aged the stellar populations using an
evolutionary correction to the luminosities and average surface
brightnesses of the galaxies, which we argue represents the maximum
amount of fading and incorporates a reasonable reddening estimate
(see text for details).
\label{fj}}
\end{figure*}
%\vskip 5mm
%%%%%%%%%%%%%%%%%%%%%%%%%%%%%%%%%%%%%%%%%%%%%%%%%%%%%%%%%%%%%%%%%%%%%
%\noindent

Our goal is to use the fundamental plane to determine the structural
family (and thus formation history) of the AGN host galaxies.
Although there is a rich morphological diversity across late-type
galaxies in the luminosity range of our sample, conceptually we wish
to distinguish between two basic formation scenarios.  On the one
hand, the structural similarity of elliptical galaxies and classical
bulges presumably reflect a violent merger history (e.g.,~Burstein
\etal\ 1997; Robertson \etal\ 2006).  On the other hand, more
quiescent evolution probably results in disk-like galaxies, either
containing pseudobulges or (when the gas and outer regions have been
removed) with the structure of spheroidal galaxies (e.g.,~Kormendy \&
Kennicutt 2004; Kormendy \etal\ 2008).  Figure 2{\it a} beautifully
illustrates the discriminatory power provided by projections of the
fundamental plane.  The projections are defined by a very clean sample
of elliptical and spheroidal galaxies in the Virgo cluster,
representing the most careful and modern photometric measurements
available from the recent study of Kormendy \etal\ (2008).  While the
positions of pseudobulges are less certain, Carollo (1999) shows that
pseudobulges also may have a lower surface brightness at a given size
than classical bulges.

Using our structural measurements, and assuming $V-I=1.34$ mag (from
Fukugita \etal\ 1995, to match the Sa colors of the galaxies, see
below), we place our targets in the fundamental plane (Fig. 2, {\it
open blue symbols}).  We plot only the bulge components of the fit,
comprising the entire galaxy in the case of the blobs, but only the
compact central component in the case of the disk galaxies.  We
exclude the bulge upper limits from Figure 2 since the upper limits on
the bulges are derived assuming that they obey the Kormendy (1977)
relation between surface brightness and effective radius.  Most
likely, given the luminosities and morphological types of these
systems, as well as the upper limits on their bulge-like components,
these seven systems contain pseudo- rather than classical bulges.

We see first of all that while our galaxies are fainter than $L^*$
($I^*=-21.5$ mag at $z=0.1$ from Blanton \etal\ 2003b and assuming
$i\arcmin - I = 0.63$ mag), they are typically a magnitude brighter
than the tip of the spheroidal sequence.  In terms of structures, we
see that while some of the blobs obey the fundamental plane of
classical bulges (GH04, GH09, GH10, and GH16), others are
significantly offset toward the locus of spheroidal galaxies (GH07,
GH12, and GH17).  The detected bulges, with the exception of GH14,
have structures more similar to spheroidal galaxies, as expected for
pseudobulges (e.g.,~Carollo 1999).  POX 52, while faint enough to
overlap with the spheroidal sequence, is more centrally concentrated
than typical for a spheroidal galaxy, as supported by the high \sers\
index, $n=4.3$.  At the same time, POX 52 also shows evidence for a
spatially resolved central component in the $B$-band (Thornton
\etal\ 2008).  Such a component would preferentially boost the central
surface brightness and decrease the effective radius, but when the
young component fades it may move onto the spheroidal sequence.  We
see tentative evidence for a similar component in GH04.  Thus the
photometric projections alone are rather ambiguous.  Let us turn now
to the stellar velocity dispersions, which provide additional
constraints on the nature of these galaxies.

In Figure 3{\it a} we present the Faber-Jackson (1976) relation for
inactive elliptical galaxies from Bernardi \etal\ (2003), along with
the locations of two spheroidal samples (Geha \etal\ 2002;
Matkovi{\'c} \& Guzm{\'a}n 2005).  Taken at face value virtually none
of our galaxies obeys the Faber-Jackson relation of elliptical
galaxies, but there are a couple of significant complications to the
interpretation.  First of all, it is unclear whether low-luminosity
ellipticals have a linear Faber-Jackson relation, as illustrated
schematically in Figure 3 (see Kormendy \& Kennicutt 2004).  There may
be a downturn in \sigmastar\ at low luminosities, making the
spheroidal, pseudobulge, and elliptical sequences overlap in this
regime.  At the same time, we cannot be sure that the dispersion
measurements are dominated by stars in the bulge component.  If there
is a bright nuclear star cluster or significant luminosity from a
face-on disk, light from these stars will artificially lower our
luminosity-weighted stellar velocity dispersion measurement.  Indeed,
the significant offset in the Faber-Jackson relation for the sample as
a whole, independent of position in the photometric projections, is
somewhat worrisome.  An additional factor that will move objects in
the Faber-Jackson plane, of course, is the mix of stellar populations.
In the following section we estimate the impact of stellar populations
and aperture size on our measurements.

\subsection{Impact of Stellar Evolution, Dust Reddening, 
and Stellar Kinematics}

%%%%%%%%%%%%%%%%%%%%%%%%%%%%%%%%%%%%%%%%%%%%%%%%%%%%%%%%%%%%%%%%%%%%
%%BoundingBox: 
%\begin{figure}
%\vbox{ 
%\vskip -0.1truein
%\hskip 0.in
\psfig{file=cmdparv3.epsi,width=0.45\textwidth,keepaspectratio=true,angle=0}
%}
\vskip -0mm
\figcaption[]{
Host galaxy location in a color-magnitude diagram.  The colors are
derived parametrically but agree with the non-parametric colors to
within $0.01 \pm 0.3$ mag.  We use the entire galaxy (not just the
bulge), but distinguish between blob ({\it circles}) and disk ({\it
triangles}) galaxies.  The relatively luminous and red galaxy is GH09.
Representative error bars are shown for the blobs and disks
respectively in the upper right-hand corner.  For reference, the
typical colors of E/S0, Sa and Scd galaxies are shown ({\it dotted};
Fukugita \etal\ 1995), and the spread in those colors are shown as
error bars on the left-hand side of the plot.  We also show the $B-I$
color of spheroidal galaxies as estimated from Fig. 8 of Gavazzi
\etal\ (2005) for systems with typical $M_B\approx-14$ mag.
\label{cmd}}
%\end{figure}
\vskip 5mm
%%%%%%%%%%%%%%%%%%%%%%%%%%%%%%%%%%%%%%%%%%%%%%%%%%%%%%%%%%%%%%%%%%%%%
%\noindent

We first estimate the impact of young stellar populations on the
observed luminosities.  Because of relatively recent or ongoing star
formation, the luminosities of our galaxies cannot be compared directly with
elliptical galaxies, which formed the majority of their stars at high
redshift (e.g.,~Thomas \etal\ 2005).  A truly fair comparison would
require knowledge of their star formation histories and metallicity
distributions, but full spectral synthesis modeling (e.g.,~Tinsley
1978) is difficult with only a single ($B-I$) color\footnote{Although
in principle additional information may be extracted from the ESI
spectra, the difficulties in modeling the combination of an AGN and
mixed stellar populations with the additional complications of
imperfect flux calibration and slit losses make such an analysis
beyond the scope of the current work.}.  In particular, without
near-infrared measurements, we do not have a strong handle on the
magnitude of extinction, which is clearly playing a significant role.
Figure 4 shows the observed location of our targets in the
color-magnitude relation, along with the fiducial colors of galaxies
of a variety of Hubble types (Fukugita \etal\ 1995).  We can see that,
with the exception of GH09, the galaxies are blue compared to
elliptical galaxies.  However, they are actually redder than we might
have expected.  For instance, GH05, GH08, and GH13 are late-type
spiral galaxies with bulge-to-total luminosity ratios $B/T < 0.07$ and
yet their $B-I$ colors are redder than that of a typical,
bulge-dominated Sa galaxy, presumably because of reddening.

In the absence of more detailed information, we use a very simplistic
prescription to account in an average way for the combined impacts of
reddening and stellar evolution on the observed color.  For all
galaxies in the sample we simply {\it assume} a single intrinsic color
of $B-I=1.67$ mag.  This color is typical of Scd galaxies (Fukugita
\etal\ 1995) and is close to the typical color of spheroidal galaxies
in the Gavazzi \etal\ sample, $B-I \approx 1.8$ mag.  From this one
assumption, we can then derive both the reddening and (with the
further assumption of a single stellar population) the total stellar
mass.  While we are not adequately recovering the star formation 
history of individual galaxies in our sample, this simple thought 
experiment provides an estimate of the impact of young 
stellar populations on the ensemble.  As we will show below, more 
sophisticated modeling yields similar final results.

Given our assumed intrinsic color, we first derive the reddening for
each galaxy assuming $A_B/A_V=1.32$ and $A_I/A_V=0.59$, as derived by
Schlegel \etal\ (1998) using the reddening curves of O'Donnell (1994;
optical) and Cardelli \etal\ (1989; ultraviolet and near-infrared).
We find typical $A_I$ values of $\langle A_I \rangle = 0.29 \pm 0.15$
mag, and have an estimate for the intrinsic luminosity of the system.
Now, for the purpose of comparison, we ask what mass-to-light ratio
(\mli) and stellar age are associated with a simple stellar population
(SSP) with a color of $B-I=1.67$ mag.  The synthesis model, from
Bruzual \& Charlot (2003), assumes a Chabrier (2003) initial mass
function (IMF), solar metallicity, and uses the Padova 1994 stellar
evolutionary tracks (Alongi \etal\ 1993; Bressan \etal\ 1993; Fagotto
\etal\ 1994a,b; Girardi \etal\ 1996).  The SSP has a color of
$B-I=1.67$ mag at an age of $2 \times 10^9$ yrs and \mli=0.6.  The
decrease in luminosity associated with an increase to \mli$=3.7$ (the
final \mli\ for this model at an age of $2 \times 10^{10}$~yrs)
corresponds to $\Delta m_I = 1.9$ mag.  Combining the effects of
reddening and stellar aging yields an absolute $I$-band magnitude and an
average surface brightness for each target, corresponding to a typical
magnitude change of $\langle \Delta m_I \rangle = 1.6 \pm 0.2$ mag. 
Using the inferred reddening and measured luminosity, and assuming an
SSP, we not only estimate the magnitude of aging, we also infer a
total stellar mass for each system.

Crucially, our approach is designed to yield an upper limit to the
impact of stellar aging.  This is because undoubtedly the galaxies
contain a significant evolved population that boosts the current
mass-to-light ratio.  For instance, {\it kcorrect} (Blanton \etal\
2003a; Blanton \& Roweis 2007) models a wide range of star formation
histories, metallicities, and reddening values, and returns \mli\
values of 1.1--1.4 for the galaxies (using either the \hst\ color, or
the SDSS magnitudes corrected for the nuclear point source using the
\hst\ $B$-band measurement), implying fading of $\Delta m_I=1-1.3$
mag.  Furthermore, particularly for the blobs, the assumed intrinsic
color may be too blue; a color of $1.8-2$ mag may be more appropriate
(e.g.,~Geha \etal\ 2002; Gavazzi \etal\ 2005)\footnote{Additional
complication is certainly added by the spatial distribution of the
star formation.  One might expect nuclear star formation associated
with AGN activity, and perhaps we are seeing that in the cases of
GH04 and GH12.  On the other hand, in the spiral galaxies the star
formation may occur preferentially in the outskirts.}.  

The stellar kinematics also impact the observed Faber-Jackson
relation.  Although we wish to know the integrated velocity dispersion
of stars in a dynamically hot stellar component within an effective
radius (\sigmastar), what we measure (\sigmam) is the
luminosity-weighted mean dispersion of all of the stars in our
aperture.  If the stars are rotation-dominated (e.g.,~in a nuclear
disk) then we will measure broadening due to unresolved rotation.  If
the light is dominated by stars in a face-on disk, then \sigmam\ will
underestimate the true dispersion.  Furthermore, our spectroscopy is
not performed out to the effective radius, but rather within a
0\farcs7 slit corresponding roughly to $r_e/2$ for most of the sample
(ranging from 20\%--90\% of $r_e$; Barth \etal\ 2005).  Apart from
broadening due to unresolved rotation, we might either systematically
over- or underestimate the true dispersion depending on the velocity
dispersion profile.  

%%%%%%%%%%%%%%%%%%%%%%%%%%%%%%%%%%%%%%%%%%%%%%%%%%%%%%%%%%%%%%%%%%%%
%%BoundingBox: 
%\begin{figure*}
%\vbox{ 
%\vskip -0.1truein
\hskip -0.1in
\psfig{file=fpkorcv5.epsi,width=0.45\textwidth,keepaspectratio=true,angle=0}
%}
%\vskip -0mm
\figcaption[]{
A reproduction of the fundamental plane projections shown in Figure 
2 with our simplistic estimates for the stellar population fading 
applied.  
\label{masshist}}
%\end{figure*}
\vskip 5mm
%%%%%%%%%%%%%%%%%%%%%%%%%%%%%%%%%%%%%%%%%%%%%%%%%%%%%%%%%%%%%%%%%%%%%
\noindent
At large radius, radial \sigmastar\ profiles are quite flat for
elliptical galaxies (e.g.,~J{\o}rgenson \etal\ 1995) and spiral bulges
(e.g.,~Pizzella \etal\ 2005).  At $r \ll r_e$, however, the profiles
are more diverse.  For instance, in the SAURON sample, the
luminosity-weighted mean stellar velocity dispersion at $0.1 r_e$ is
$\sim 20\%$ higher than at the effective radius (Cappellari \etal\
2006).  In spheroidal galaxies, conversely, \sigmastar-drops of
$>30\%$ are commonly observed (e.g.,~Geha \etal\ 2002; van Zee \etal\
2004; De Rijcke \etal\ 2004, 2006).  On the other hand, at $r \approx
0.5 r_e$, the impact of these excursions is actually minor.  The
Cappellari \etal\ profiles show that \sigmastar\ at $0.3 r_e$ is
already within $10\%$ of \sigmastar\ at $r_e$.  Likewise, the Geha
\etal\ profiles have already flattened at a few tenths of $r_e$.

Empirically, if the \sigmam\ values are depressed due to a drop toward
the center, then sources that we observed at the lowest fraction of
$r_e$ should have the largest deviations from the Faber-Jackson
relation.  Even including the disks with only bulge upper limits,
which in general lie closer to the Faber-Jackson relation, we find no
evidence for a correlation between $r/r_e$ and Faber-Jackson offset; a
Kendall's $\tau$ test indicates a $70\%$ probability that no
correlation is present.  Alternatively, the galaxies with the most
prominent blue centers (and therefore the strongest nuclear weighting
of a possible cold component) may show the largest deviations from
Faber-Jackson.  However, we see no significant correlation between
either the integrated parametric color or an inner color, with a $30
\%$ chance of no correlation.  The inner color is measured from the
raw data between the radius where the model is more than twice the
luminosity of the AGN and the effective radius of the bulge (or disk
in the case of upper limit systems).

Although we see no obvious trends between offsets in the Faber-Jackson
plane and either our aperture or the galaxy color, we now make simple
estimates of the relative contribution to \sigmam\ from a non-bulge
component.  We begin with a hypothetical nuclear stellar concentration
in the blob systems by assigning $50\%$ of the AGN luminosity to a
nuclear star cluster with a velocity dispersion that is $60\%$ of the
assumed \sigmastar\ at larger radius.  We then compute the
luminosity-weighted dispersion within our aperture using the measured
light profile and assuming a constant \sigmastar\ outside of the
unresolved nucleus.  In general the contamination to \sigmastar\ is
small, but \sigmam\ can be depressed by as much as $10\%$ compared to
the assumed true \sigmastar.

In particular for the disk systems, we worry that \sigmam\ may be
dominated by unresolved rotation.  Again, a disk component may
increase or decrease \sigmam\ relative to \sigmastar, depending on
whether the disk is closer to face-on or edge-on.  Our images show
that the disk-dominated galaxies are rather close to face-on, and so
it is worth investigating the magnitude of \sigmastar\ depression that
a kinematically cold disk component can cause, using our photometric
decompositions.  We calculate a one-dimensional luminosity-weighted
stellar velocity dispersion, assuming a single value for the bulge
component and 10~\kms\ for the disk (Bottema 1993).  For the systems
with detected bulges, the resulting contamination is small
(5\%--14\%), whereas for those with bulge upper limits, the disk
completely dominates the light at nearly all radii, and thus the
corrections can be as large as 80\%.  At the same time, of course,
rotation is likely to contribute significantly to \sigmam\ in these
systems.  Finally, we perform the same test with the blobs, using the
disk limits derived in \S 4.1.  In this case the corrections are very
small (<2\%), since by definition we do not detect the disk
significantly over the radii of our observations.

There is an additional factor that may impact the \sigmastar\
measurements, which is the presence of the AGN.  As emphasized,
e.g.,~by Greene \& Ho (2006a), both narrow forbidden emission lines
and broad \feii\ pseudo-continuum can significantly bias \sigmastar,
particularly in the \mgb\ spectral region.  However, there are a
variety of reasons to believe that these effects are minor in the case
of this sample.  The first is that for the $\sim 50\%$ of the sample
with calcium triplet measurements, Barth \etal\ (2005) find good
agreement with the \mgb\ dispersions.  Another is that the high
resolution of the ESI data mitigates the contamination from emission
lines.  Finally, in general the sources are not strong \feii\
emitters.  Altogether, then, we suspect that the \sigmastar\
measurements are relatively free of bias from AGN contamination.  In
short, for the majority of cases we expect stellar contributions from
a disk or compact component to depress \sigmastar\ at the $10\%$
level, while for the three bulge upper limit systems, the dispersion
is probably dominated by the disk component.

We now have crude estimates for the impact of both stellar population
effects and stellar kinematics on the observed location of objects in
the fundamental plane.  We have attempted to derive the largest
possible corrections, particularly in the case of stellar population
aging, so that between the observations and our corrected values we
bracket reality for all systems.  Correspondingly, the maximum
depression to \sigmastar\ that we derived for any of the blobs was $10
\%$.  In this spirit, and for schematic purposes only, Figure 3{\it b}
shows the Faber-Jackson relation with the estimated stellar population
evolution and the \sigmastar\ measurements boosted by $10\%$ for the
blobs.  We also show the fundamental plane projections after fading in
Figure 5.

%%%%%%%%%%%%%%%%%%%%%%%%%%%%%%%%%%%%%%%%%%%%%%%%%%%%%%%%%%%%%%%%%%%%
%%BoundingBox: 
\begin{figure*}
\vbox{ 
\vskip -3.4mm
\hskip +0.95in
\psfig{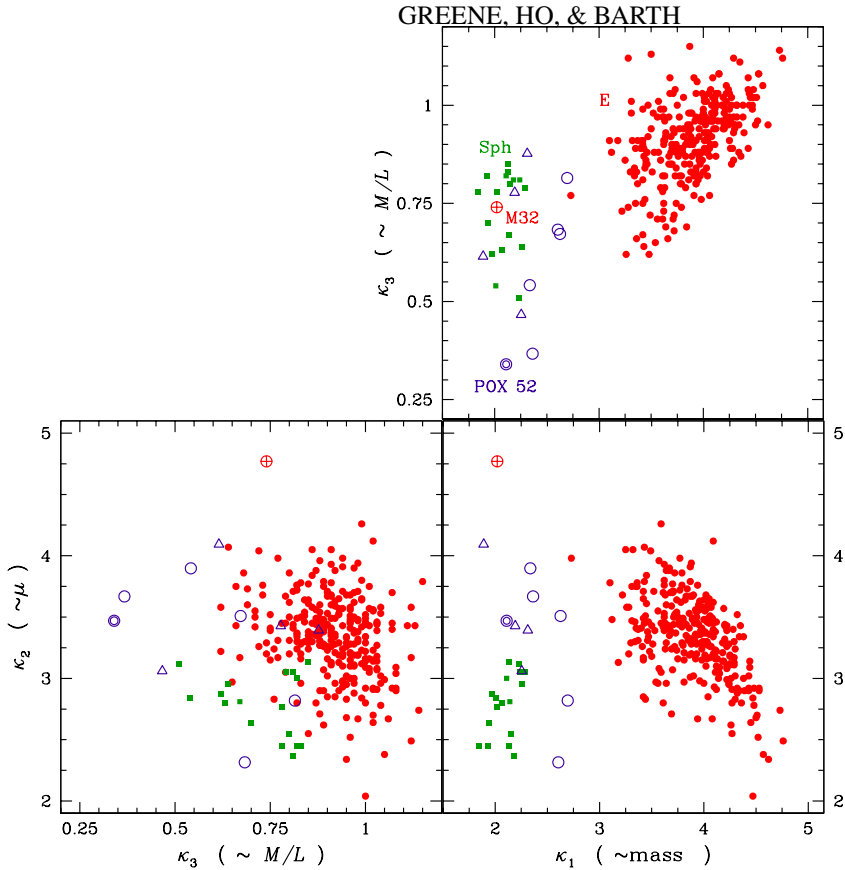}
}
\vskip -0mm
\figcaption[]{
$\kappa$-space projections of the fundamental plane.  Symbols as in
Figure 2, with comparison data taken from the compilation of Bender
\etal\ (1992) and Geha \etal\ (2002).  We plot the observed data with
no corrections here.  In this space, our sample is clearly distinct
from the locus of elliptical galaxies, mainly driven by the offsets in
the Faber-Jackson plane.
\label{masshist}}
\end{figure*}
%\vskip 5mm
%%%%%%%%%%%%%%%%%%%%%%%%%%%%%%%%%%%%%%%%%%%%%%%%%%%%%%%%%%%%%%%%%%%%%
%\noindent

What can we conclude then about the nature of the blobs?  Based on the
fundamental plane of Figure 2 alone, three of them (GH07, GH12, and
GH17) look more like spheroidal galaxies that are abnormally bright
due to ongoing star formation, while the other four are ambiguous.
Their photometric structures look like elliptical galaxies, but their
dispersions place them below the Faber-Jackson relation.  It could be,
as suggested by Kormendy \& Kennicutt (2004), that Faber-Jackson is
non-linear for low-luminosity elliptical galaxies, but much better
statistics are needed to determine whether the effect is due to
ongoing star formation or structural differences.  We have shown that
with reasonable assumptions about young stellar populations, the
systems fade onto the spheroidal locus (Fig. 5).  While GH09 is both
too luminous and too red to be anything but an elliptical galaxy, we
suspect that GH04, GH10, and GH16 will fade onto the spheroidal
sequence.  Thornton \etal\ (2008) draw similar conclusions about the
nearby galaxy POX 52.  While POX 52 has a high \sers\ index of 4.3, it
is not exactly a scaled-down giant elliptical galaxy like M32; its low
velocity dispersion of 36 \kms\ places it closer to the sequence of
spheroidal galaxies (Figs. 3, 6).

To further emphasize this latter point, and as a convenient means of
summarizing the fundamental plane results, we also present the
``$\kappa$-space'' formalism in Figure 6 (e.g.,~Bender \etal\ 1992).
This is a coordinate transformation in which the axes are proportional
to galaxy mass, surface brightness, and mass-to-light ratio,
respectively (only galaxies with \sigmastar\ measurements can be
included).  It is clear that, due primarily to the offset in
Faber-Jackson, our sample galaxies are significantly offset from the
locus of elliptical galaxies in $\kappa$-space.  In this formalism the
blobs separate from the elliptical sequence in a similar fashion as
inactive spheroidal galaxies.

%%%%%%%%%%%%%%%%%%%%%%%%%%%%%%%%%%%%%%%%%%%%%%%%%%%%%%%%%%%%%%%%%%%%
%%BoundingBox: 
\begin{figure*}
\vbox{ 
\vskip -0.5truein
\hskip -0.1in
\psfig{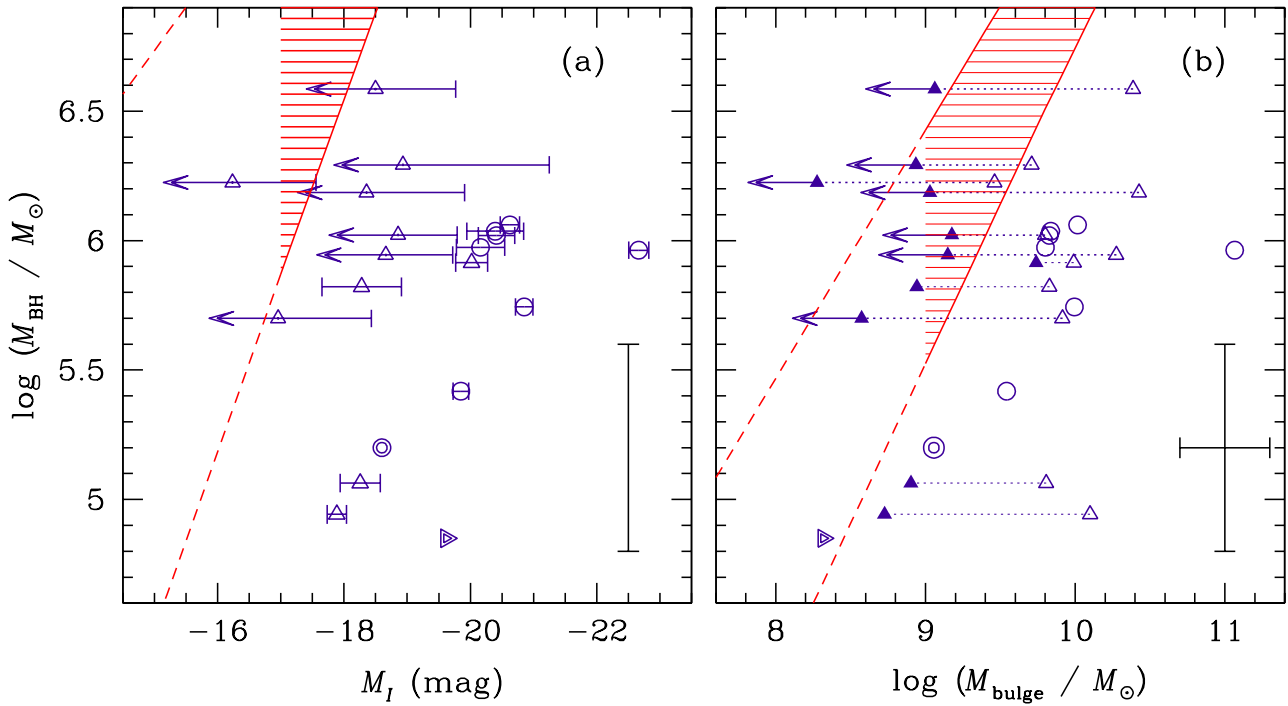}
}
%\vskip -0mm
\figcaption[]{
({\it a}) \mbh\ as a function of bulge luminosity in the $I$ band.
Different host morphologies are noted with different symbols as in
Fig. 2 above.  Upper limits to bulge-like components are shown as
left-pointing arrows. A 0.4 dex uncertainty on \mbh\ (Greene \& Ho
2006b) is shown schematically in the lower right-hand corner.  Also
included are the well-studied low-mass BHs in POX 52 ({\it double
circle}) and NGC 4395 ({\it double triangle}).  The hatched region
shows the fit from Tundo \etal\ (2007) to inactive BHs with dynamical
mass estimates, shifted to $I$ band assuming a color $r\arcmin - I =
1.07$ mag for E/S0 galaxies from Fukugita \etal\ (1995).  ({\it b})
\mbh\ as a function of bulge mass as estimated for a single stellar
population from Bruzual \& Charlot (2003; see text for details).
Total masses are shown as open circles for blobs and open triangles
for disks, while bulge masses alone (derived assuming a constant \mli)
are filled triangles.  The \mbh-$M_{\rm bulge}$ relation of H{\"a}ring
\& Rix (2004) is shown as the shaded region.  The blobs are offset
from the expected relation by $\sim 1$ dex, while the bulge components
appear to be rather consistent with the relation at higher mass.
Overall, we see that the scatter is much larger at low mass.  Although
the stellar masses are subject to a number of systematic uncertainties
discussed in the text, for a wide range of reasonable assumptions the
observed offset remains.  The representative error in the stellar
masses of 0.3 dex is due to systematic effects as discussed in
Kannappan \& Gawiser (2007), but note that in practice our stellar
masses should be lower limits rather than true measurements.
\label{masshist}}
\end{figure*}
%\vskip 5mm
%%%%%%%%%%%%%%%%%%%%%%%%%%%%%%%%%%%%%%%%%%%%%%%%%%%%%%%%%%%%%%%%%%%%%

\subsection{Relations with BH Mass}

We have identified a population of BHs in galaxies without classical
bulges, strongly confirming that formation of the latter is not a
prerequisite for existence of the former.  Nevertheless, given the
different structural properties of the host galaxies, it is
interesting to investigate whether the relationship between \mbh\ and
$L_{\rm bulge}$ is preserved for this sample.  The virial BH masses are
directly compared with both the bulge luminosities (Fig. 7{\it a}) and
bulge masses (Fig. 7{\it b}).  These spheroidal and pseudobulge
systems are striking outliers in the \mlb\ plane, in the sense that
the galaxies are overluminous compared to the low-mass extrapolation
of the \mlb\ relation.  All of the blobs and detected bulges are
offset independent of their location in the fundamental plane.

The observed offset is not an artifact of improper AGN-host galaxy 
decomposition.  The AGN accounts for at most a few tenths
of the total luminosity, and thus cannot be a major contributor to
uncertainties in the galaxy magnitudes.  It is, however, worth
considering the possibility that the BH masses are systematically
biased to low values.  In the case of the blobs, if we wished to make
the galaxies obey the \mlb\ relation the BH masses would need to be
biased by 2 orders of magnitude on average.  In principle it is
possible that individual objects in the sample have considerably
larger BH masses than suggested by their virial masses (as would
occur, for instance, if the BLR were a face-on disk).  We have shown,
however, that the radiative properties of the sample are consistent
with expectations for systems radiating at a high fraction of their
Eddington limit [based on their optical (Greene \& Ho 2004), radio
(Greene \etal\ 2006), and X-ray (Greene \& Ho 2007a) properties].
Thus, it is highly unlikely that these systems are actually, on
average, radiating at $<1\%$ of their Eddington luminosities.

As with the fundamental plane, we speculate that the observed offset
is due to young stellar populations rather than a true excess of mass,
since the real correlation is presumably with bulge mass rather than
luminosity (e.g.,~H{\"a}ring \& Rix 2004).  The fiducial stellar
population assumptions described above provide estimates of the host
stellar masses (Fig. 7{\it b}).  Again, we argue that the corrections
are maximal since we have completely neglected an old (faint) stellar
component.  Therefore, these masses represent a lower limit on the
true stellar mass.  Even so, all of the detected bulges remain
systematically offset from the relation toward higher stellar mass.
Reassuringly, our masses agree remarkably well with the {\it kcorrect}
masses, $\langle {\rm log}~M_{\rm our} - {\rm log}~M_{\rm
kcor} \rangle = 0.02 \pm 0.06$, which include a more
sophisticated treatment of both the star formation history and the
dust reddening.  However, the treatment of stellar evolution and the
adopted IMF are the same for the two methods, and systematic
differences in these two assumptions dominate uncertainties in the
mass estimates (e.g.,~Maraston 2005; Kannappan \& Gawiser 2007).
Nevertheless, the BH masses for the detected bulges and blobs are an
order of magnitude lower than expected from the inactive \mgal\
relation, which is a large offset compared to the probable systematics. Taking
the virial masses at face value, we find that the ratio of BH mass to
bulge mass is considerably lower in these galaxies compared to their
high-mass cousins.  Interestingly, G1 (Gebhardt \etal\ 2002, 2005) and
Omega Cen (Noyola \etal\ 2008) display the opposite trend (their ratios of
BH mass to bulge mass are closer to $10^2$ than to $10^3$).

Finally, we recall that our sample as a whole is biased toward the
host galaxies that are luminous enough to be targeted
spectroscopically by the SDSS survey (Greene \& Ho 2007b).  We may
simply have selected the luminous tail of the full distribution, and
we do see that the bulges of the disk galaxies fall closer to the
inactive relation than do the blobs (Fig. 7{\it b}).  While we cannot
quantify the true distribution of host galaxy mass at a given \mbh, we
can rule out that these galaxies are drawn from the same distribution
as the inactive sample.  There is a 1 mag scatter ($1~\sigma$) about a
fixed BH mass in the inactive sample (where we have used the best-fit
line of Tundo \etal\ 2007 and the data compiled in H{\"a}ring \& Rix
2004).  At the median \mbh\ of the blobs ($\sim 10^6$~\msun), the
median host luminosity is $M_I = -20.4$ mag, which is $2.4~\sigma$
from the expected central value of $M_I=-17.9$ mag.  Thus, in order to
be consistent with the observed central value and scatter for the
inactive relation, we would require that the galaxy luminosity
function rise by a factor of 300 between $M_I=-18$ mag and $M_I -20$
mag, which can be ruled out categorically (e.g.,~Blanton \etal\
2003b).  Of course, we must implicitly assume that the active fraction
is constant over this range.  While AGN activity may depend on mass
and Hubble type at some level, the dependence cannot change so
radically.  Furthermore, we have ignored the additional scatter
introduced by the virial BH mass measurements, largely because that
scatter is poorly quantified.  However, if we adopt the 0.4 dex
scatter measured in the \msigma\ relation by Greene \& Ho (2006b), the
calculation is unchanged.  Although we cannot measure the true
distribution of galaxy luminosity around BH mass, we can robustly
conclude that it is different in this regime.  Finally, we cannot
presently rule out that some combination of systematic errors in the
BH masses, selection bias in the host population, and uncertainties in
the host galaxy masses are conspiring to cause the observed offset.
While we view the possibility of a large systematic offset in the BH
masses of the blob galaxies as unlikely, only a
reverberation mapping campaign for these galaxies can help to settle
this potential caveat.

\section{Discussion}

While the absence of a BH in M33 (Gebhardt \etal\ 2001) tells us that
not all late-type galaxies contain BHs, this sample confirms
unambiguously that galaxies without classical bulges may nevertheless
contain nuclear BHs.  We already knew of a few such systems.  The
prototype is the BH in the late-type spiral galaxy NGC 4395, but AGNs
have been detected recently in two other late-type spiral galaxies,
NGC 3621 (Satyapal \etal\ 2007) and NGC 1042 (Shields \etal\ 2008).
There is also a BH in POX 52, which is probably a spheroidal (Barth
\etal\ 2004; Thornton \etal\ 2008).  The galaxies
presented here, as a whole, span a wide range in morphological type,
but in general are inconsistent with being classical bulges.  Adopting
reasonable assumptions about the stellar populations of the galaxies
in this study, we find that the compact blobs deviate from the
fundamental plane of ellipticals toward the locus of spheroidal
galaxies.  We argue that the disk galaxies are inconsistent with
having classical bulges, although we do not have direct structural
measurements in most cases.  Interestingly, other samples of low-mass,
highly accreting BHs are also found preferentially in late-type spiral
galaxies with a high fraction of strong bars (e.g.,~Crenshaw \etal\
2003, although see also Ohta \etal\ 2007) and nuclear dust spirals
(Deo \etal\ 2006).

Apparently galaxies can form central BHs even when they do not form
classical bulges, but the process that results in the \mgal\ relation
in bulge-dominated systems may or may not operate in the low-mass
regime.  Recall that, unfortunately, observational constraints from
dynamical BH masses are very sparse in this regime.  There are no
dynamical BH mass measurements in spheroidal galaxies (aside from the
upper limit in NGC 205 [Valluri \etal\ 2005] and very loose upper
limits for a handful of Virgo spheroidals [Geha \etal\ 2002]), and so
a different \mgal\ relationship, or a significant increase in the
scatter, cannot be ruled out for galaxies of this type.  There are a
handful of dynamical masses measured in pseudobulges, which may
suggest that the BHs are in general undermassive compared to $M_{\rm
bulge}$ (C.~Y.~Peng \& L.~C.~Ho 2008, in preparation) or \sigmastar\ (Hu
2008).  However, some of the masses are based on highly uncertain
maser kinematics, and it may be premature to draw strong
conclusions. It is clear that the \mgal\ relation breaks down at low
mass, at least for this sample.  We have only weak constraints on the
relation between BH and pseudobulge mass, but the blob galaxies are
clearly overly massive compared to their central BHs.  Before we
attempt to interpret this result in the context of the galaxy
population as a whole, however, we must consider the possibility that
our result is peculiar to actively accreting BHs.

The BHs in our sample are currently radiating and thus gaining mass.
Many authors have suggested that vigorously accreting local BHs are
growing toward the \msigma\ relation, based on estimates for
\sigmastar\ from narrow emission-line widths (typically the
\oiii~$\lambda 5007$ line; e.g.,~Grupe \& Mathur 2004; Bian \& Zhao
2004).  On the other hand, the reliability of the \oiii\ linewidth,
which often displays a strongly asymmetric profile or is highly
blueshifted, has been questioned (e.g.,~Boroson 2005; Greene \& Ho
2005a), and indeed when lower-ionization lines are used instead, the
deviations from the \msigma\ relation appear to be mitigated
(e.g.,~Komossa \& Xu 2007).  Although it is observationally quite
challenging to obtain direct \sigmastar\ measurements for BHs in a
high accretion state in general, those galaxies for which it has been
possible do not appear to deviate from the inactive \msigma\ relation
(Barth \etal\ 2005; Botte \etal\ 2005).

An interesting and independent constraint on the galaxy potential may
be derived from circular velocity measurements.  There is some
suggestion that, at a given \mbh, $v_{\rm max}$ as measured from
\ion{H}{1} is lower at high Eddington ratio, and the trend is
significantly strengthened when one considers the galaxy dynamical
mass (Ho \etal\ 2008).  The sense is the same as the offset of $\sim
0.17$~dex seen by Onken \etal\ (2004) and Greene \& Ho (2006b) in the
\msigma\ relation of active galaxies.  Furthermore, M.~Kim \etal\ (in
preparation) find that the most highly accreting local broad-line AGNs
lie systematically below the \mlb\ relation.  In all of these cases,
however, the measured offset is closer to a factor of 2 in BH mass,
consistent with a single Eddington-limited growth phase over a
Salpeter time.  The order-of-magnitude offset that we see here is much
harder to reconcile with the inactive relationship.  It seems more likely
that the \mgal\ relation is simply different in this regime.

As an aside, we also note that the apparent evolution in the \mgal\
relation at high redshift is in the opposite sense to that observed
here.  At redshifts ranging from $0.37 < z < 6.4$ numerous authors
have found evidence for evolution in BH-host scaling relations, but
the ratio of BH to bulge mass appears to increase at earlier times
(e.g.,~Peng \etal\ 2006a,b; Woo \etal\ 2006; Shields \etal\ 2006;
Salviander \etal\ 2006; Ho 2007). If we were seeing the impact of
continuing evolution in \mgal, then we must either discount all of the
above studies (although see Lauer \etal\ 2007b) or once again posit
that mechanisms relating BH mass and galaxy mass are different for
low-mass systems.  Even if the \mgal\ relation continues to evolve to
a small extent in low-mass active systems at low redshift, our data
suggest that the final relation looks different at low mass.

Given that low-mass galaxies have different, presumably more
quiescent, formation histories, a different relationship with their
central BHs is not necessarily surprising.  In fact, the BHs are of
low enough mass that they are consistent with forming by a variety of
seed production mechanisms (e.g.,~Portegies Zwart \& McMillan 2002;
Koushiappas \etal\ 2004; Begelman \etal\ 2006) and then growing
minimally ever since.  Perhaps in the absence of the violent, gas-rich
events thought to be responsible for the formation of elliptical
galaxies and classical bulges, BHs do not necessarily grow
significantly.  We have no direct constraints on the growth history of
low-mass BHs from existing observations of the active BH mass
function, which unfortunately are not constraining at low BH mass and
high redshift (e.g.,~Heckman \etal\ 2004; Kollmeier \etal\ 2006;
Greene \& Ho 2007b; Shen \etal\ 2008b).  If BH growth at low mass is
driven by more stochastic processes (e.g.,~Hopkins \& Hernquist 2006),
then we might indeed expect to see a bias toward more massive host
galaxies at a given BH mass.

On the other hand, while the \mgal\ relation appears to change with
mass, we do have some evidence for continuity in the \msigma\ relation
for these systems (Barth \etal\ 2005).  Given that the galaxies do not
obey the fundamental plane of elliptical galaxies and classical
bulges, we would not expect them to simultaneously obey \msigma\ and
\mgal.  However, they could have easily obeyed neither\footnote{Barth
\etal\ (2005) and Greene \& Ho (2006b) both note the apparent
flattening at low mass in the \msigma\ relation slope.  Even adopting
the maximum contamination of $20\%$ to \sigmastar, then, the ensemble
would still be consistent with the \msigma\ relation.  Note that the
data presented in Barth \etal\ (2005) have been used to support
possible non-linearity in the \msigma\ relation (e.g.,~Wyithe 2006).
It is our opinion that, given the outstanding uncertainties in both
the BH mass scale and the contributions of rotation to some objects,
these claims are rather premature.  }.  Better statistics are needed,
particularly since a fraction of our measurements are probably
dominated by rotation, but it is at least intriguing that the \msigma\
relation seems to be preserved, even for the massive globular clusters
G1 and Omega Cen (Gebhardt \etal\ 2002, 2005; Noyola \etal\ 2008).

Are there any proposed mechanisms for establishing BH-bulge relations
that specifically preserve an \msigma\ relation but not necessarily
\mgal?  Explanations for the \msigma\ or \mgal\ relation that rely on
BH self-regulation by construction do not predict any mass or
structural dependence in the final relation.  The BH grows until it is
massive enough to unbind the gas in the galaxy (e.g.,~Silk \& Rees
1998; Murray \etal\ 2005).  One alternative is suggested by Peng
(2007), who proposes that merging in a hierarchical universe, combined
with a mass function that rises toward low mass, will lead to a tight
correlation between BH mass and bulge mass at the luminous end.  In
this scenario the BH-galaxy relations at low mass begin to depend in
detail on the merger histories of the galaxies, and it is not clear
that an \msigma\ relation would necessarily result.  Perhaps the most
promising alternative is that BH growth really depends on a
(relatively) local stellar velocity dispersion, as would be expected
for instance in a model where the accretion disk is fed by the
capture of individual stars (Miralda-Escud{\'e} \& Kollmeier 2005).

We should note that there are other regimes in which the \msigma\ and
\mgal\ relations cannot be simultaneously valid.  The most luminous
elliptical galaxies, particularly brightest cluster galaxies, also
deviate from the fundamental plane of regular ellipticals and
classical bulges; while \sigmastar\ saturates at $\sim 400$~\kms\ (and
correspondingly \mbh$\approx 3 \times 10^9$~\msun), the luminosities
and effective radii of brightest cluster galaxies continue to rise
(e.g.,~Bernardi \etal\ 2007; Lauer \etal\ 2007a).  Although it is
difficult to obtain dynamical BH masses for brightest cluster galaxies
outside of the Virgo Cluster, Lauer \etal\ argue that in this regime
bulge luminosity may be a better predictor of \mbh\ than is
\sigmastar.  Massive elliptical galaxies are characterized by a
deficit of light in their inner profiles, perhaps because the
innermost stars have been ejected through three-body interactions with
a binary BH (e.g.,~Faber \etal\ 1997).  Within this framework, the
size of the core serves as an indirect probe of \mbh.  Since core size
correlates more strongly with bulge luminosity than with \sigmastar,
Lauer \etal\ (2007a) propose that bulge luminosity is a better
predictor of BH mass.  More dynamical BH masses are needed, at both
mass extremes, to settle this question.

In short, it is clear that the relationship between BH mass and
``bulge'' mass changes in low-mass systems without classical bulges.
It is implausible that the BHs will grow sufficiently to lie on the
\mgal\ relation.  Given the different formation histories of the
low-mass galaxies compared to classical bulges and elliptical
galaxies, it seems more likely that the \mgal\ relation just changes
in this regime.  In contrast, the \msigma\ relation appears to
continue to low mass with less scatter than the \mgal\ relation.  It
will be interesting to see whether these trends persist with larger
and less biased samples.

\section{Summary}

We have used \hst-ACS images, in combination with \sigmastar\
measurements, to examine the host galaxy structures for a sample of
low-mass active galaxies.  The systems in this study were selected
based on BH mass rather than galaxy properties, in order to
investigate the range of host structures and luminosities that host
low-mass BHs.  We find that the host galaxies are relatively faint;
with $\langle M_I \rangle =-20.7$ mag, they are $\sim 1$ mag
below $L^*$ at $z=0.1$.  The colors are typical of Sa galaxies,
$\langle B-I \rangle = 2.0 \pm 0.2 $ mag.  There are 11
disk galaxies (although two of those have ambiguous
classifications) and 7 compact blobs with no evidence of a
disk component.  All of the disk galaxies are consistent with
containing a bulge, but in seven of the 11 disks we can only place
limits on the bulge luminosities.

The structural measurements from the ACS images, combined with
\sigmastar\ measurements from Barth \etal\ (2005), allow an
investigation of the fundamental plane locations for the bulge-like
components of the galaxies.  We find that once stellar population and
reddening effects are accounted for, the compact blobs are closer to
the locus of spheroidal galaxies than that of elliptical galaxies or
classical bulges.  We argue that the disk galaxies probably all
contain pseudobulges, although we detect them in only a handful of
cases.  Our result provides strong confirmation that BHs are found
even in the absence of classical bulges.  At the same time, the ratio
of BH to bulge mass in these systems is considerably lower than in
more massive inactive systems.  Perhaps, therefore, the process that
builds bulges is required to stimulate healthy BH growth.
Alternatively, it may be that the BH is more sensitive to the central
stellar velocity dispersion than to the details of galaxy formation.

Some immediate progress will be made in clarifying the trends we
observe.  We are currently in the process of collecting comparable
data for nearly an order of magnitude more objects (Greene \& Ho
2007c).  Hopefully it will be possible to obtain reverberation mapping
masses for low-mass systems in addition to NGC 4395 (Peterson \etal\
2005).  Optical monitoring campaigns, to identify suitable candidates,
are underway.  Both of these endeavors will provide a clearer picture
of the relation between BH mass and galaxy properties at low mass.
Meanwhile, alternate search techniques (e.g.,~Satyapal \etal\ 2007;
Ulvestad \etal\ 2007; Desroches \& Ho 2008) 
may help to mitigate the biases in host galaxy
properties suffered by this optically selected sample.  At the same
time, an increased sample of dynamical BH masses in spiral galaxies
would be extremely useful in elucidating the true BH-galaxy relations
at low mass.

\acknowledgements

We thank J.~Kormendy both for many useful conversations, and for
providing the fundamental plane data prior to publication.  It would
not be possible to thank C.~Y.~Peng adequately, not only for
developing and supporting GALFIT, but also for answering countless
technical questions that arose during the course of this research, and
for stimulating science discussions as well. We thank P.~Martini for
providing the cosmic ray removal software, K.~Gebhardt for helpful
suggestions and A.~Socrates for interesting discussions.  Finally, we
acknowledge the anonymous referee for a very careful reading of the
manuscript and comments that significantly improved this paper.
Support for J.~E.~G. was provided by NASA through Hubble Fellowship
grant HF-01196, and support for L.~C.~H was provided by
HST-GO-10586.01-A, both awarded by the Space Telescope Science
Institute, which is operated by the Association of Universities for
Research in Astronomy, Inc., for NASA, under contract NAS 5-26555.
Research by A.~J.~B. is supported by the National Science Foundation
under Grant No. AST-0548198.  Funding for the SDSS has been provided
by the Alfred P. Sloan Foundation, the Participating Institutions, the
National Science Foundation, the U.S. Department of Energy, the
National Aeronautics and Space Administration, the Japanese
Monbukagakusho, the Max Planck Society, and the Higher Education
Funding Council for England. The SDSS web site is {\tt
http://www.sdss.org/}.  This research has made use of data obtained
from the High Energy Astrophysics Science Archive Research Center
(HEASARC), provided by NASA's Goddard Space Flight Center.

\appendix

\section{Notes on Individual Galaxies}

%%%%%%%%%%%%%%%%%%%%%%%%%%%%%%%%%%%%%%%%%%%%%%%%%%%%%%%%%%%%%%%%%%%%
%%BoundingBox: 
\begin{figure*}
\vbox{ 
\vskip -0.1truein
\hskip +0.75in
\psfig{file=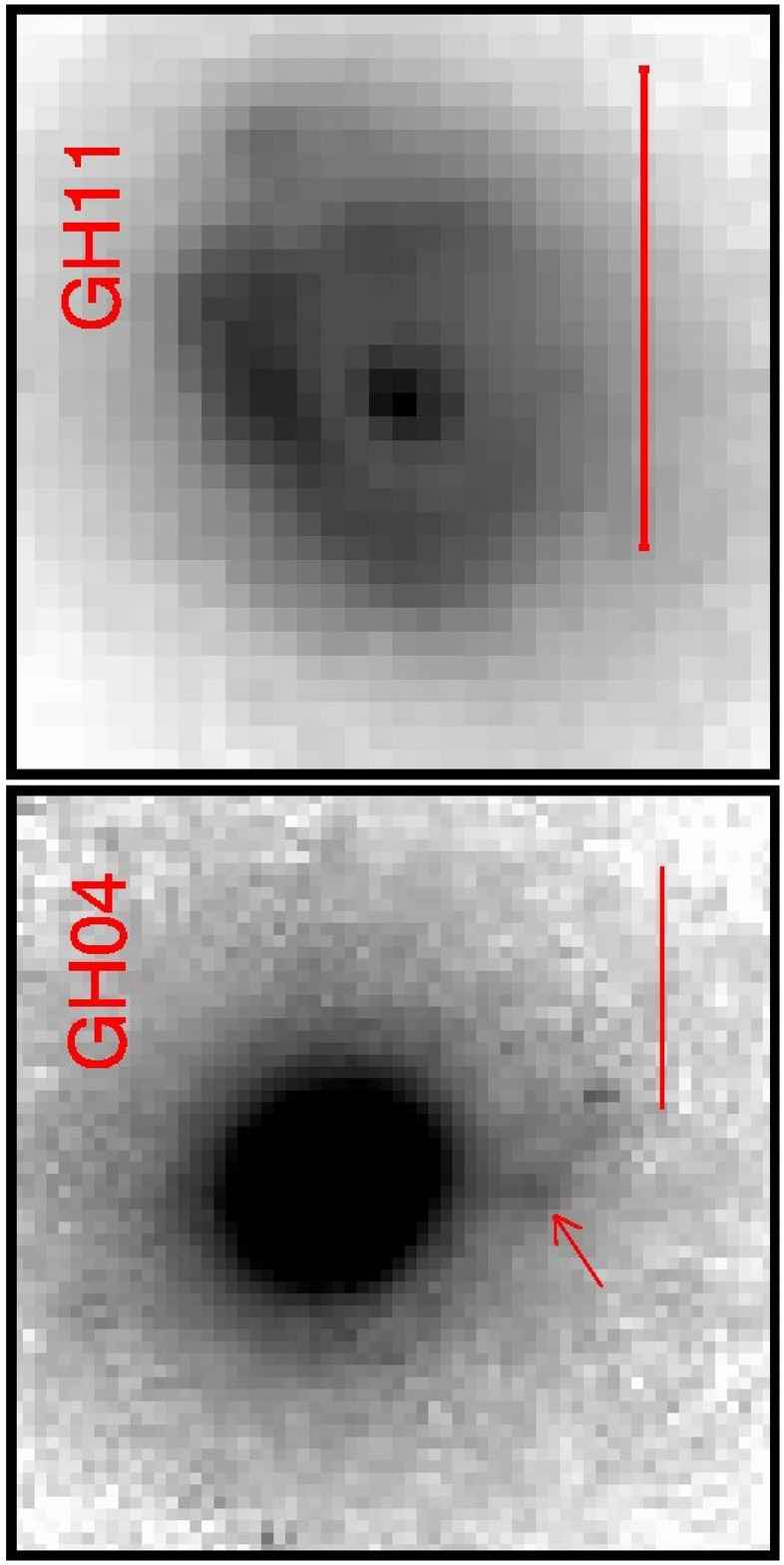,width=0.35\textwidth,keepaspectratio=true,angle=-90}
}
%\vskip -0mm
\figcaption[]{
$B$-band images for GH04 and GH11.  Scale bars are 1\arcsec\ in length.
Note the tail in GH04 and the blue knot or ring in GH11.
\label{masshist}}
\end{figure*}
\vskip 5mm
%%%%%%%%%%%%%%%%%%%%%%%%%%%%%%%%%%%%%%%%%%%%%%%%%%%%%%%%%%%%%%%%%%%%%

\subsection{GH04}

In the $I$ band GH04 is a compact bulge-like system.
Structurally it sits on the fundamental plane, although it is
significantly offset in the Faber-Jackson relation.  The most
interesting aspect of this system is the ``tail'' most easily seen in
the $B$-band image (Fig. A8), extending to the East of the nucleus.
Given that we have strong upper limits on the radio emission from GH04
(Greene \etal\ 2006), this feature is more likely associated with the
accretion of a small satellite or star formation in the host galaxy
than some sort of jet feature.

\subsection{GH07}

GH07 is one of the three clear spheroidals in the sample.  It is
unremarkable in terms of luminosity and color, but is one of the most
asymmetric objects in the sample.  The residuals from the GALFIT in
Figure 1 show this asymmetry most clearly. There is a clear excess
above the symmetric model to the North-West.  There is also a faint
galaxy in the field to the North.

\subsection{GH09}

GH09 is the most luminous and the reddest galaxy in our sample.  It
very clearly lies on the fundamental plane of ellipticals and
classical bulges (although we do not have a stellar velocity
dispersion measurement for it).  In terms of the properties of the
objects in our sample, this galaxy is a clear outlier.  It would be
very interesting if this massive galaxy actually hosts a $10^6$~\msun\
BH, but in this case the more likely scenario seems to be that we have
seriously underestimated the BH mass.  We could benefit from a higher
signal-to-noise ratio spectrum at higher resolution to perform an
improved fit to the linewidth, but it would also be desirable to
obtain reverberation mapping for this source, to see if it really is
such a deviant system.

\subsection{GH11}

This galaxy has the dubious distinction of being the most difficult to
model.  Although it is very compact, it appears to have a distinct
shelf of emission in the inner few arcseconds (Figs. 1, A8).  In
Figure A5 we show the blue image of GH11, in which an asymmetric ring
is very evident.  It appears that the galaxy has an asymmetric ring of
star formation with a the radius is 0\farcs3, or $\sim 500$ pc away
from the nucleus, which is consistent with the sizes of nuclear rings
in mid-type spiral galaxies (e.g.,~Buta \& Combes 1996).  In support
of the idea that this ring of excess emission is actively forming
stars is the fact that GH11 is the bluest galaxy in our sample, and
also (interestingly) has the largest \oii/\oiii\ ratio in the sample.
Ho (2005; see also Kim \etal\ 2006) suggests that the \oii/\oiii\ may
be a good indicator of ongoing star formation activity in AGNs, since
\oii\ is intrinsically weak in high-ionization Seyfert galaxies.

\end{document}